\documentclass[aps,pre,reprint,groupedaddress,superscriptaddress]{revtex4-1}

\usepackage[utf8]{inputenc}
\usepackage{amsmath}
\usepackage{amssymb}
\usepackage{bbold}
\usepackage{bm}
\usepackage{graphicx}

\usepackage{titlesec}
\usepackage{xcolor}
\definecolor{ioplink}{cmyk}{0.82,0.28,0,0.18} 
\usepackage{hyperref}
\hypersetup{
  colorlinks=true,
  linkcolor=ioplink,  
  citecolor=ioplink,  
  urlcolor=ioplink,   
  filecolor=ioplink
}
\usepackage{cleveref}
\usepackage{orcidlink}

\renewcommand{\thesection}{\arabic{section}}
\renewcommand{\thesubsection}{\thesection.\arabic{subsection}}
\makeatletter
\renewcommand{\p@subsection}{}
\makeatother
\titleformat{\section}[block]{\normalfont\large\bfseries}{\thesection.}{0.5em}{}
\titleformat{\subsection}[block]{\normalfont\normalsize\bfseries}{\thesubsection.}{0.5em}{}

\begin{document}

\title{Supermoir\'{e} reconstruction and domain-resolved valley topology in helical multilayer graphene}

\author{Kyungjin Shin\,\orcidlink{0009-0001-2368-0729}}
\affiliation{Department of Physics and Astronomy, Seoul National University, Seoul 08826, Korea}

\author{Nicolas Leconte\,\orcidlink{0000-0002-1209-9656}}
\altaffiliation{Current affiliation: Catalan Institute of Nanoscience and Nanotechnology (ICN2), CSIC and BIST, Campus UAB, Bellaterra, 08193 Barcelona, Spain}
\affiliation{Department of Physics, University of Seoul, Seoul 02504, Korea}

\author{Jeil Jung\,\orcidlink{0000-0003-2523-0905}}
\email[Corresponding author: ]{jeiljung@uos.ac.kr}
\affiliation{Department of Physics, University of Seoul, Seoul 02504, Korea}

\author{Hongki Min\,\orcidlink{0000-0001-5043-2432}}
\email[Corresponding author: ]{hmin@snu.ac.kr}
\affiliation{Department of Physics and Astronomy, Seoul National University, Seoul 08826, Korea}

\date{\today}

\begin{abstract}
\noindent
Extending moir\'{e} graphene beyond twisted bilayers, helical trilayer graphene has emerged as a supermoir\'{e} platform hosting topological bands and correlated electronic states.
Here we establish a domain-resolved framework for helical multilayer graphene that links supermoir\'{e} relaxation to low-energy electronic structure and valley topology.
Real-space lattice calculations show that relaxation reconstructs the system into locally periodic single-moir\'{e} domains, enabling a local continuum description.
Within each domain, downfolding the first-shell model yields effective Hamiltonians near the Dirac points that reveal how the low-energy spectrum decomposes into folded Dirac sectors.
The resulting valley Chern topology is domain-dependent and includes gate-tunable responses, as independently corroborated by real-space lattice calculations.
Our results identify folded Dirac sectors as the organizing principle of thicker helical graphene stacks, with local stacking families determining the band character and valley topology.  
\end{abstract}

\maketitle

\section{Introduction}
\label{Sec:Sec1}

\noindent
The discovery of superconductivity and correlated insulating phases~\cite{Cao2018a,Cao2018b} arising from flat electronic bands~\cite{Li2010,Laissardiere2010,Morell2010,Bistritzer2011,Tarnopolsky2019,Utama2021,Lisi2021} in magic-angle twisted bilayer graphene (t2G) positioned moir\'{e} superlattices as a setting for strongly correlated quantum matter.
This advance launched twistronics~\cite{Carr2017,Balents2020,Andrei2021,Hennighausen2021}, in which quantum phases are engineered by tuning the twist angle, material composition, and layer number, as well as through electrostatic gating in van der Waals heterostructures.
Beyond t2G, twistronics has been extended to a wide range of systems,
including alternating-twist multilayer graphene~\cite{Khalaf2019,JPark2021,Hao2021,JPark2022,Zhang2022,Nguyen2022,JShin2021,Leconte2022ATMG,KShin2023,Leconte2024,KShin2024},
rhombohedral multilayer graphene aligned with hexagonal boron nitride~\cite{Chen2019,Chen2020,Chen2022,Chittari2019,Galeano2021,YPark2023,Han2024,Lu2024,Dong2024,Zheng2025,Liu2026},
and twisted transition metal dichalcogenides~\cite{Naik2018,Vitale2021,Devakul2021,Cai2023,Zeng2023,HPark2023,Crepel2023,Reddy2023,Xu2024,Wang2024,Duran2024,YXia2025,Guo2025,Zhu2025,Tuo2025}.

Among these architectures, helical trilayer graphene (h3G), a structure in which three consecutive layers are twisted by identical angles~\cite{Mora2019,Zhu2020,Mao2023}, has emerged as a promising moir\'{e} platform for topological and correlated physics~\cite{LXia2025,Bocarsly2026,Guerci2024PRR,Nakatsuji2023,Devakul2023}.
In h3G, interference between the two adjacent moir\'{e} patterns generates a longer-period modulation known as the supermoir\'{e} (or moir\'{e}-of-moir\'{e}) pattern.
Despite this structural complexity, h3G exhibits a locally periodic single-moir\'{e} structure in which supermoir\'{e} reconstruction induces a relative displacement between the two moir\'{e} superlattices~\cite{Nakatsuji2023,Devakul2023,Hoke2026}.
This structural simplification reduces the multiscale problem to an effective continuum description organized by local stacking geometry.

The non-interacting band structure of h3G features an isolated set of low-energy bands near charge neutrality.
These bands become strongly flattened and are separated from the remote bands by a sizable gap on the order of $100$~meV at the magic angle $\theta \approx 1.8^{\circ}$~\cite{Guerci2024PRR,Devakul2023,Hoke2026,Guerci2024PRB,Yang2024,Popov2023PRR}, analogous to the isolated flat bands in magic-angle t2G.
In addition to band flattening, h3G introduces an emergent domain degree of freedom in its topological landscape,
going beyond the valley, spin, and sublattice characterizations commonly used for moir\'{e} graphene systems~\cite{Tarnopolsky2019,Bultinck2020PRL,Bultinck2020PRX}.
Specifically, each domain carries a distinct total Chern number determined by its local stacking registry, giving rise to a Chern mosaic pattern across the supermoir\'{e} lattice~\cite{Guerci2024PRR,Datta2024,Kwan2024,Kwan2025,LXia2025}.

A natural question in this context is how moir\'{e} physics extends to helical multilayers, where additional layers can qualitatively reshape the band structure and topology.
To address this question, we develop a domain-resolved continuum description of helical $N$-layer graphene (h$N$G), supported by real-space lattice calculations.
We show that supermoir\'{e} relaxation reorganizes helical multilayers into local single-moir\'{e} domains, within which the low-energy bands decompose into folded Dirac sectors, each formed by collecting the monolayer Dirac cones folded onto a given moir\'{e} Brillouin-zone (mBZ) momentum.

The remainder of this paper is organized as follows.
In Sec.~\ref{Sec:Sec2}, we introduce our model for h$N$G, including the supermoir\'{e} lattice construction and the continuum Hamiltonian, together with a perturbative downfolding procedure.
In Sec.~\ref{Sec:Sec3}, we explore the electronic structure and valley topology of h3G and h4G, focusing on the $\alpha\beta$ and $\alpha\beta\alpha$ domains, respectively.
We then generalize our findings for this alternating sequence to an arbitrary layer number $N$,
while other stacking families, such as $\alpha\alpha\alpha$ and $\alpha\beta\gamma$, are detailed in the Appendix.
Finally, Sec.~\ref{Sec:Sec4} presents the discussion and concluding remarks.

\section{Model}
\label{Sec:Sec2}

\subsection{Supermoir\'{e} lattice}
\label{subSec:2A}

\noindent
We consider h$N$G, consisting of vertically stacked graphene layers with a consecutive twist angle $\theta$ between adjacent layers.
Building on recent works~\cite{Nakatsuji2023,Devakul2023}, we model h$N$G directly on the supermoir\'{e} length scale using real-space lattice calculations.
This approach requires the construction of commensurate supercells that reproduce,
within a chosen tolerance, both the target twist angle and the equilibrium lattice constant of graphene.
We construct such commensurate cells following the standard prescription~\cite{Hermann2012,Leconte2022TBG} and its multilayer extension~\cite{Leconte2024}, which allows a single-moir\'{e} treatment of multilayers with multiple moir\'{e} patterns.

The commensurate supercell for h$N$G is uniquely specified by $2N$ integers
\(
(m_1,n_1,m_2,n_2,\ldots,m_N,n_N),
\)
which define a set of integer transformation matrices
\begin{equation}
{\bm M}_{\ell} =
\begin{pmatrix}
n_{\ell} & m_{\ell} \\
- m_{\ell} & n_{\ell} + m_{\ell}
\end{pmatrix},
\qquad \ell = 1,\ldots,N.
\label{eq:indice_general}
\end{equation}
Here, ${\bm M}_{\ell}$ corresponds to layer $\mathrm{L}_{\ell}$, with $\mathrm{L}_{1}$ taken as the untwisted reference layer.
The supercell lattice vectors $\bm{r}_{1,2}$ are common to all layers and are related to the primitive lattice vectors of each layer through
\begin{equation}
\begin{pmatrix}
\bm{r}_{1} \\
\bm{r}_{2}
\end{pmatrix}
=
{\bm M}_{\ell}
\begin{pmatrix}
\bm{a}_1^{(\ell)} \\
\bm{a}_2^{(\ell)}
\end{pmatrix}.
\label{eq:Hermann_general}
\end{equation}
This condition ensures the commensurability of all $N$ layers within a single supercell.
The lattice mismatch and twist angle of layer $\mathrm{L}_{\ell}$ relative to $\mathrm{L}_{1}$ are given by 

\begin{subequations}
\label{eq:lattice_general}
\begin{align}
\alpha_{1\ell}
&= \frac{|\bm{a}_1^{(1)}|}{|\bm{a}_1^{(\ell)}|}
= \sqrt{\frac{n_{\ell}^{2} + m_{\ell}^{2} + n_{\ell} m_{\ell}}{n_{1}^{2} + m_{1}^{2} + n_{1} m_{1}}},
\label{eq:lattice_general_a}\\
\theta_{1\ell}
&= \theta_\ell - \theta_1 \notag\\
&= \cos^{-1}\!\left[
\frac{
2 n_{1} n_{\ell} + 2 m_{1} m_{\ell} + n_{1} m_{\ell} + m_{1} n_{\ell}
}{
2 \alpha_{1\ell} (n_{1}^{2} + m_{1}^{2} + n_{1} m_{1})
}
\right].
\label{eq:lattice_general_b}
\end{align}
\end{subequations}

\begin{figure}[t!]
\includegraphics[width=1.0\linewidth]{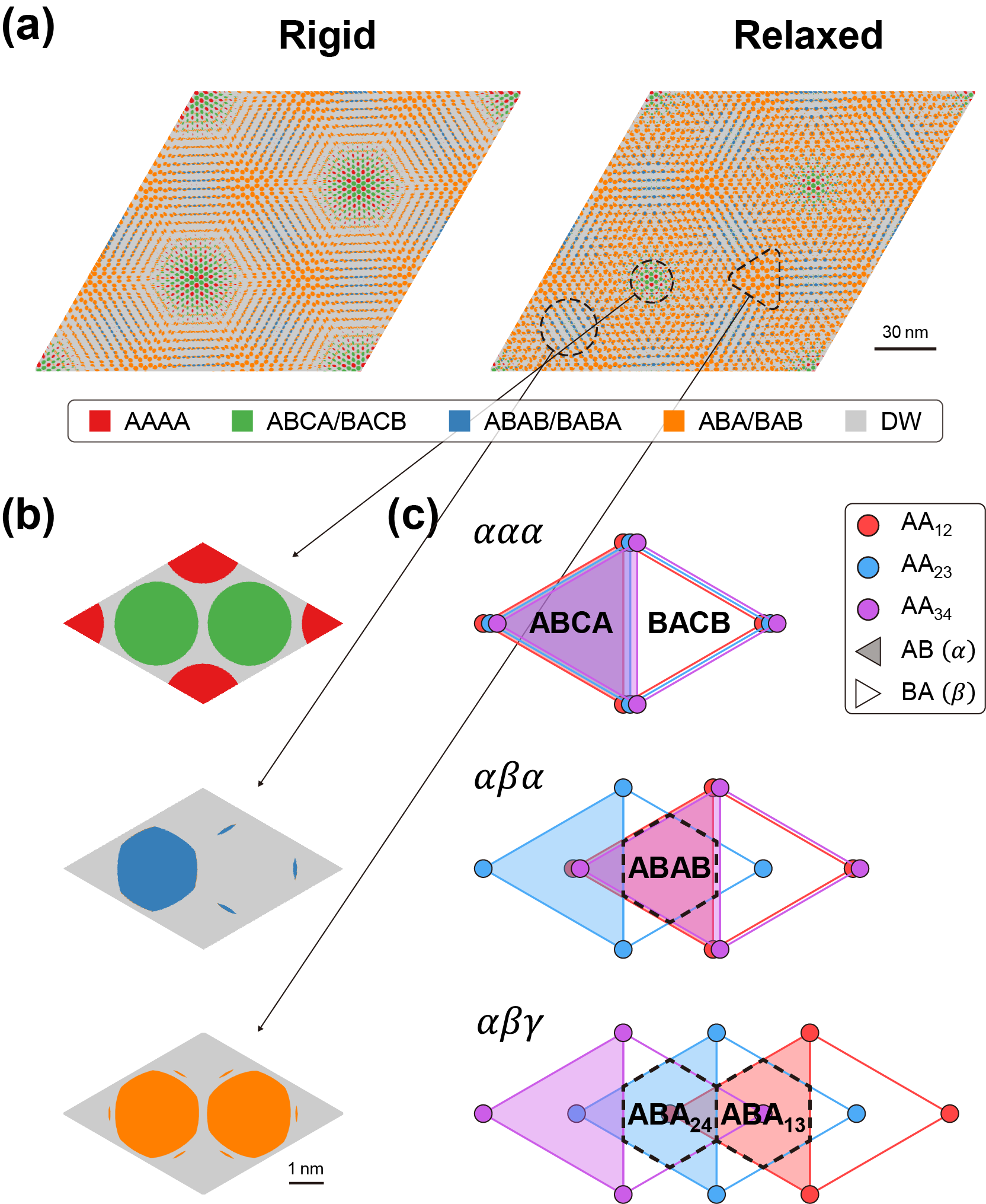}
\caption{
\textbf{Supermoir\'{e} lattice reconstruction and local single-moir\'{e} domain formation in h4G.}
(a) Supermoir\'{e} stacking maps of h4G for the rigid (unrelaxed, left) and relaxed (right) structures at a consecutive twist angle of $\theta = 3^\circ$.
Grey regions denote domain walls, where the local stacking does not belong to a well-defined high-symmetry class.
(b) Representative local stacking distributions corresponding to the distinct single-moir\'{e} domains.
(c) Geometric schematics of the idealized $\alpha\alpha\alpha$ (top), $\alpha\beta\alpha$ (middle), and $\alpha\beta\gamma$ (bottom) domains, illustrating the relative displacement between adjacent moir\'{e} patterns and the resulting local high-symmetry stackings.
Scale bars, 30 nm in (a) and 1 nm in (b).
}
\label{fig:fig1}
\end{figure}

For h3G, h4G, and h5G considered in this work, the specific integer tuples $(m_{\ell},n_{\ell})$, as well as the resulting lattice constants and twist angles, are summarized in Appendix~\ref{Sec:AppendixA} (Table~\ref{tab:MasterSupercell}).
The resulting structures are fully relaxed under fixed-cell periodic boundary conditions using \texttt{LAMMPS}~\cite{Plimpton1995}. Interlayer interactions are described by the registry-dependent DRIP potential~\cite{Wen2018}, whose parameters are informed by EXX-RPA calculations~\cite{Leconte2017}.
To account for the energetic preference of Bernal over rhombohedral stacking in multilayer graphene, we additionally include a next-nearest layer Gaussian correction~\cite{Park2025},
\begin{equation}
V_{2\mathrm{nn}}(r) = -V \exp\!\left[-b \left(r^2 - z^2\right)\right],
\end{equation}
where $r$ is the three-dimensional interatomic distance and $z=6.626$~\AA\ is the fixed out-of-plane separation between next-nearest layers.
We use $V = 1.0$~meV and $b = 3.0$~\AA$^{-2}$, which, together with DRIP, reproduces the experimentally relevant Bernal--rhombohedral energy difference of $0.06$~meV/atom.
Intralayer C--C interactions are modeled using the second-generation REBO2 potential~\cite{Brenner2002}.
Structural relaxation is carried out using the FIRE energy minimization algorithm~\cite{Bitzek2006FIRE} with a time step of $0.001$~ps and a force-convergence criterion of $10^{-3}$~eV/\AA.

In graphene moir\'{e} systems, lattice relaxation reconstructs the superlattice by enlarging energetically favorable AB/BA (Bernal) regions and separating them by domain walls~\cite{Jung2014,Uchida2014,Wijk2015,Chebrolu2019,Leconte2022TBG}.
We label the AB and BA local stacking domains by $\alpha$ and $\beta$, respectively.
The supermoir\'{e} relaxation in h3G forms locally commensurate $\alpha\beta$/$\beta\alpha$ domains while shrinking $\alpha\alpha$ domains, reminiscent of moir\'{e}-scale relaxation in t2G.

Adding one more layer produces a richer supermoir\'{e} tessellation.
As shown in Fig.~\ref{fig:fig1}, h4G hosts coexisting $\alpha\alpha\alpha$, $\alpha\beta\alpha$/$\beta\alpha\beta$, and $\alpha\beta\gamma$/$\gamma\beta\alpha$ domains, corresponding to moir\'{e}-scale AA-, Bernal-, and rhombohedral-like stacking configurations, respectively.
This suggests that the moir\'{e}-of-moir\'{e} lattice resulting from the additional layer can be interpreted as a patchwork of domains defined by local stacking sequences of moir\'{e} superlattices.
We therefore use h4G as the representative case for the multilayer domain structure, since it is the minimal extension beyond h3G in which all three domain types coexist in the relaxed structure.

Before relaxation, the local stackings are nearly uniformly distributed across the h4G structure [Fig.~\ref{fig:fig1}(a), left].
These local stackings are obtained using the method outlined in Appendix~\ref{subSec:A3}.
At this stage, the stacking distribution is determined solely by the embedding of local configurations within the larger stacking domains.
Explicitly, the $\alpha\alpha\alpha$ domains comprise one AAAA region and two ABCA/BACB regions;
the $\alpha\beta\alpha$ domains contain one ABAB region together with two surrounding domain-wall (DW) regions;
and the $\alpha\beta\gamma$ domains consist of two distinct three-layer ABA regions,
$\mathrm{ABA}_{13}$ and $\mathrm{ABA}_{24}$ (spanning layers $\mathrm{L}_{1}$--$\mathrm{L}_{3}$ and $\mathrm{L}_{2}$--$\mathrm{L}_{4}$, respectively),
while the remaining layer either continues the Bernal sequence or introduces a stacking fault.

After relaxation, significant lattice reconstruction occurs [Fig.~\ref{fig:fig1}(a), right].
The energetically unfavorable AAAA regions shrink markedly, which in turn reduces the embedded ABCA/BACB regions within the same $\alpha\alpha\alpha$ domains.
In contrast, the favorable ABAB/BABA Bernal regions expand, increasing the area fraction of the $\alpha\beta\alpha$ domains.
Finally, the three-layer ABA regions within the $\alpha\beta\gamma$ domains are likewise sufficiently stable to expand under relaxation.

As a result, the initially continuous distribution of local stackings reorganizes into an almost perfectly patterned array of single-moir\'{e} domains.
These observations validate a local single-moir\'{e} description within each of the $\alpha\alpha\alpha$, $\alpha\beta\alpha$, and $\alpha\beta\gamma$ domains.
Accordingly, we model the low-energy electronic structure within each reconstructed domain using a single-moir\'{e} continuum Hamiltonian.

\subsection{Continuum Hamiltonian}
\label{subSec:2B}

\noindent
The continuum Hamiltonian for h$N$G in the $K$ valley reads
\begin{equation}
\label{eq:model_H}
H_{K} = 
\begin{pmatrix}
h_{1}  & T_{12}(\bm{r}) & 0 & \cdots \\
T_{12}^{\dagger}(\bm{r}) & h_{2} & T_{23}(\bm{r}) & \cdots\\
0 & T_{23}^{\dagger}(\bm{r}) & h_{3} & \cdots\\
\vdots & \vdots & \vdots & \ddots
\end{pmatrix},
\end{equation}
where $h_{\ell}=-i\hbar v_{\rm F}\bm{\sigma}\cdot\vec{\nabla}+V_{\ell}\mathbb{I}_{2}$ is the intralayer Dirac Hamiltonian of the $\ell$th layer.
The layer-dependent potentials $V_{\ell}$ describe a perpendicular electric field and are chosen such that $V_{\ell+1}-V_{\ell} = \Delta/(N-1)$, so that the total potential drop is $V_{N}-V_{1}=\Delta$.
The Fermi velocity is set to $v_{\rm F} \simeq 10^{6}$~m/s, corresponding to a nearest-neighbor intralayer hopping term $t = -3.1$~eV and lattice constant $a = 2.46$~{\AA}.

In our model, the moir\'{e} tunneling between adjacent twisted layers takes the form
\begin{gather}
T_{\ell\ell'}(\bm{r})
= \sum_{n=0,\pm} e^{-i\bm{q}_{n}\cdot(\bm{r}-\bm{d}_{\ell\ell'})}\,
T^{n}\Bigl[1-i\lambda_{\rm MDT}\,(\hat{\bm{q}}_{n,\perp}\!\cdot\!\vec{\nabla})\Bigr]
\notag \\
T^{0} = 
\begin{pmatrix}
w' & w\\
w  & w'
\end{pmatrix},\quad
T^{\pm} = 
\begin{pmatrix}
w' & w e^{\mp i2\pi/3}\\
w e^{\pm i2\pi/3} & w'
\end{pmatrix},
\label{eq:model_T}
\end{gather}
where $\bm{q}_{n}=k_{\theta}\left[\sin\!\left(\tfrac{2\pi n}{3}\right),-\cos\!\left(\tfrac{2\pi n}{3}\right)\right]$ are the tunneling wavevectors with $k_{\theta}=\frac{8\pi}{3a}\sin(\tfrac{\theta}{2})$, and
$\hat{\bm{q}}_{n,\perp}=\left[\cos\!\left(\tfrac{2\pi n}{3}\right),\sin\!\left(\tfrac{2\pi n}{3}\right)\right]$ is the unit vector perpendicular to $\bm{q}_{n}$.
We use $w'=0.0939$~eV and $w=0.12$~eV~\cite{Chebrolu2019}, with the ratio $w'/w<1$ reflecting the effect of lattice corrugation~\cite{Uchida2014,Wijk2015}.
To capture the experimentally relevant particle--hole asymmetry, we include momentum-dependent tunneling (MDT) via the leading gradient correction by setting $\lambda_{\mathrm{MDT}}=-2.3$~{\AA}~\cite{Kwan2025,Hoke2026}.
The spatial configuration of the domains [Fig.~\ref{fig:fig1}(c)] is governed by the moir\'{e} displacement vector $\bm{d}_{\ell\ell'}$.
In the main text, we focus on the $\alpha\beta\alpha$ family, characterized by $\bm{d}_{\ell\ell'} = (-1)^{\ell+1}\bm{\delta}$ with $\bm{\delta}=[2\pi/(3\sqrt{3}k_{\theta})]\hat{\bm{x}}$.
The remaining families, namely $\alpha\alpha\alpha$ $(\bm{d}_{\ell\ell'} = \mathbf{0})$ and $\alpha\beta\gamma$ $(\bm{d}_{\ell\ell'} = -2\ell\bm{\delta})$, are addressed in Appendix~\ref{Sec:AppendixB}.

When studying the Hamiltonian analytically, we adopt the first-shell model truncated to the three primary $\bm{G}$ vectors under the rigid assumption ($w = w' = 0.12$~eV) as in Ref.~\cite{KShin2023}, neglecting the MDT term ($\lambda_{\rm MDT}=0$) for simplicity.
Numerically, we incorporate atomic corrugation ($w \neq w'$) and MDT, extending the plane-wave cutoff up to the fifth moir\'{e} reciprocal shell ($G \leq 5$) to ensure spectral convergence.
Unless otherwise specified, a representative twist angle of $\theta = 3^{\circ}$ is used, slightly above the magic-angle regime ($\sim2^{\circ}$),
where the moir\'{e} bands remain moderately dispersive, providing a clearer view of the underlying electronic structure and topology.

To obtain a low-energy effective model near the moir\'{e} Dirac points, we downfold the Hamiltonian onto the low-energy subspace $P$ by perturbatively eliminating its complement $Q$~\cite{Lowdin1962}.
The projected Schr\"{o}dinger equation takes the form
\begin{equation}
\label{eq:Heff1}
\left[(\varepsilon-H_{PP})-H_{PQ}(\varepsilon-H_{QQ})^{-1}H_{QP}\right]\psi_{\mathrm{low}}=0.
\end{equation}
Expanding $(\varepsilon-H_{QQ})^{-1}$ to first order in $\varepsilon$ yields the eigenvalue equation $\mathcal{H}_{\mathrm{eff}}\psi_{\rm low}=\varepsilon\psi_{\rm low}$ with the effective Hamiltonian
\begin{equation}
\label{eq:Heff2}
\mathcal{H}_{\mathrm{eff}}
= S\left(H_{PP}-H_{PQ}H_{QQ}^{-1}H_{QP}\right)S^{\dagger},
\end{equation}
where the term in parentheses corresponds to the conventional second-order perturbative contribution, while the prefactors $S$ and $S^{\dagger}$ account for the energy-scale renormalization and satisfy $S^{\dagger}S=(1+H_{PQ}H_{QQ}^{-2}H_{QP})^{-1}$~\cite{Koshino2009,Zhang2010}.
Given $H_{PQ} \sim w$ and $H_{QQ} \sim \hbar v_{\rm F}k_\theta$, the $Q$ sector admixture to the low-energy states has a weight of order $\alpha^{2}$, where $\alpha=w/(\hbar v_{\rm F}k_\theta)$ is the dimensionless coupling.
We therefore expand the effective Hamiltonian up to fourth order in $\alpha$, a perturbative treatment that remains valid over the twist-angle range considered here.
This analytical approach establishes a domain-resolved effective model that captures the essential low-energy physics in h$N$G, as we elaborate below.

\section{Results}
\label{Sec:Sec3}

\subsection{\texorpdfstring{\protect\MakeLowercase{h}3G}{h3G}}
\label{subSec:h3G}

\begin{figure}[t!]
\includegraphics[width=1.0\linewidth]{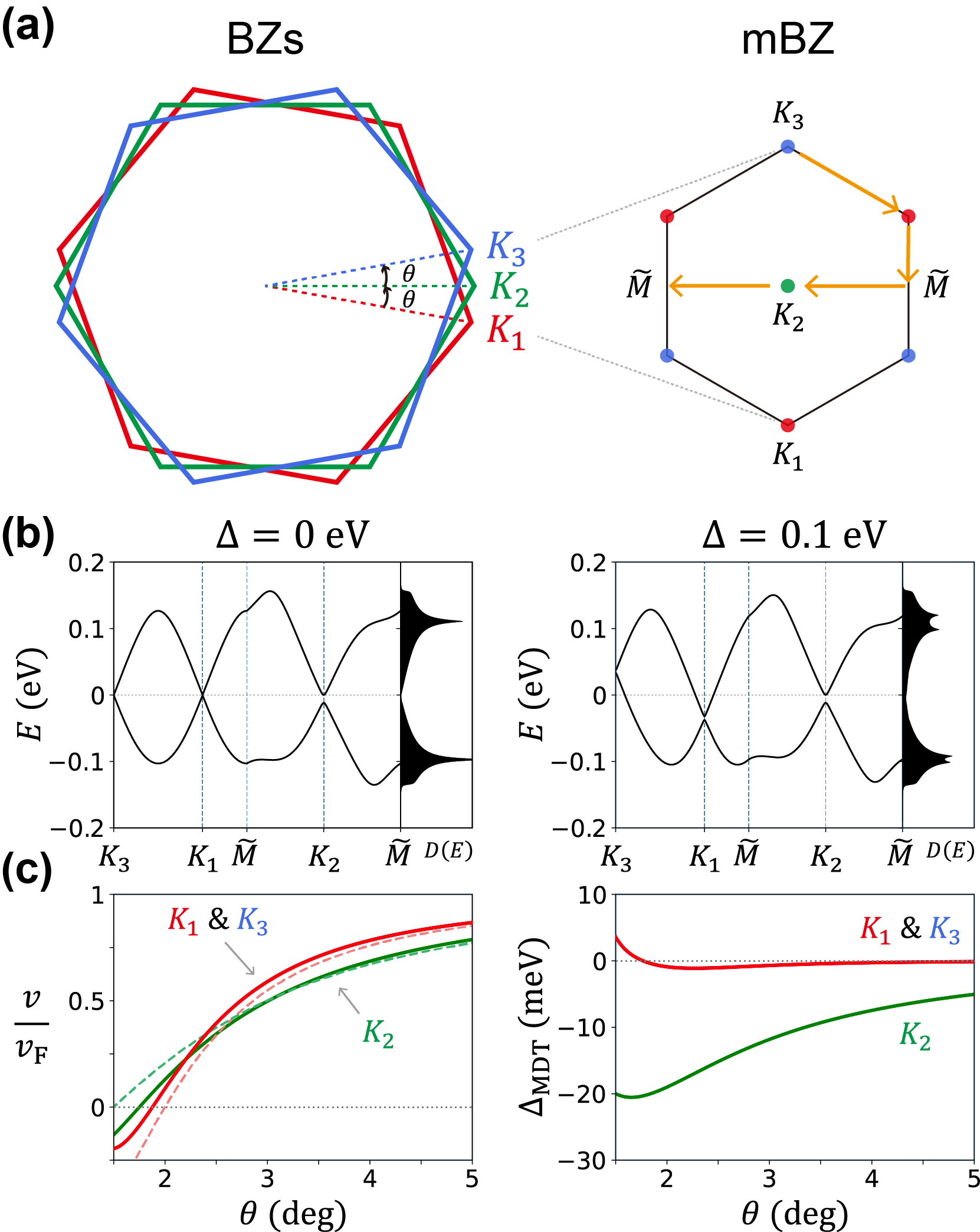}
\caption{
\textbf{Moir\'{e} geometry and low-energy electronic structure of h3G.}
(a) Brillouin zones (BZs) of the three graphene layers for a consecutive twist angle $\theta$ (left), and the corresponding mBZ with $K_{1,2,3}$ and $\tilde{M}$ indicated (right).
(b) Low-energy band structures along the high-symmetry path in the mBZ and density of states at $\theta=3^{\circ}$ for applied biases of $\Delta=0$ (left) and $\Delta=0.1$~eV (right).
(c) Twist-angle dependence of the renormalized Dirac velocities $v/v_{\rm F}$ (left) and MDT-induced gaps $\Delta_{\rm MDT}$ (right).
In the left panel, solid and dashed lines denote the full continuum calculations and the analytical first-shell model, respectively; the right panel shows the full-continuum results.
Red, green, and blue indicate $K_1$, $K_2$, and $K_3$, respectively.
}
\label{fig:fig2}
\end{figure}

\noindent
We begin with h3G, the minimal helical graphene system in which supermoir\'{e} reconstruction occurs, to introduce the domain-resolved continuum framework.
The low-energy electronic structure of h3G in the $\alpha\beta$ domain exhibits gapped Dirac cones at the folded Dirac points $K_{1,2,3}$, each associated with the Dirac point of a corresponding graphene layer [Fig.~\ref{fig:fig2}(a,b)].
To elucidate this layer-resolved structure, we downfold the first-shell Hamiltonian near each $K_{\ell}$ onto the corresponding layer $\mathrm{L}_{\ell}$, from which the low-energy Dirac state originates, while treating the other layers perturbatively.
This reduction yields an effective two-band Dirac model for each sector:
\begin{equation}
\mathcal{H}^{K_{\ell}}_{\mathrm{eff},\,\alpha
\beta}(\bm{k},\Delta)
= \hbar v^{(\ell)} (\bm{k}\cdot\bm{\sigma})
+ \Delta_{0}^{(\ell)} \mathbb{I}_{2}
+ \Delta_{z}^{(\ell)} \sigma_{z},
\end{equation}
where $v^{(\ell)}$ is the renormalized Dirac velocity at $K_{\ell}$, while $\Delta_{0}^{(\ell)}$ and $\Delta_{z}^{(\ell)}$ denote the bias-induced on-site energy shift and Dirac mass, respectively.

Near $K_{1}$, the low-energy states are localized on $\mathrm{L}_{1}$, while $\mathrm{L}_{2}$ and $\mathrm{L}_{3}$ enter as high-energy sectors.
Downfolding the first-shell Hamiltonian in this $(1+2)$-block form gives
\begin{equation}
\label{eq:H3G_K1}
\begin{aligned}
\frac{v^{(1)}}{v_{\rm F}} &= \frac{1-9\alpha^{2}-18\alpha^{4}}
{\sqrt{(1+36\alpha^{4})(1+63\alpha^{4})}}, \\
\frac{\Delta_{0}^{(1)}}{\Delta} &= -\frac{1-6\alpha^{2}+18\alpha^{4}}{2(1+36\alpha^{4})(1+63\alpha^{4})}, \\
\frac{\Delta_{z}^{(1)}}{\Delta} &= -\frac{27\alpha^{4}}{2(1+36\alpha^{4})(1+63\alpha^{4})}.
\end{aligned}
\end{equation}
Moir\'{e} tunneling therefore renormalizes the Dirac velocity, while the perpendicular electric field generates both an on-site energy shift and a Dirac mass that gaps the cone at $K_{1}$.

The structure at $K_{2}$ is qualitatively different.
Because the low-energy states are localized on the middle layer $\mathrm{L}_{2}$, they experience no net potential under a symmetric bias, and the first-order $\Delta$ terms vanish:
\begin{equation}
\label{eq:H3G_K2}
\frac{v^{(2)}}{v_{\rm F}} = \frac{1-6\alpha^{2}}{1+12\alpha^{2}}, \qquad
\frac{\Delta_{0}^{(2)}}{\Delta} = 
\frac{\Delta_{z}^{(2)}}{\Delta} = 0.
\end{equation}
The states near $K_{3}$, in turn, reside on the bottom layer $\mathrm{L}_{3}$.
Projecting out the high-energy $\mathrm{L}_{1}$ and $\mathrm{L}_{2}$ sectors results in
\begin{equation}
\label{eq:H3G_K3}
\begin{aligned}
\frac{v^{(3)}}{v_{\rm F}} &= \frac{1-9\alpha^{2}-18\alpha^{4}}
{\sqrt{(1+36\alpha^{4})(1+63\alpha^{4})}}, \\
\frac{\Delta_{0}^{(3)}}{\Delta} &= \frac{1-6\alpha^{2}+18\alpha^{4}}{2(1+36\alpha^{4})(1+63\alpha^{4})}, \\
\frac{\Delta_{z}^{(3)}}{\Delta} &= \frac{27\alpha^{4}}{2(1+36\alpha^{4})(1+63\alpha^{4})}.
\end{aligned}
\end{equation}
These bias-induced terms reverse sign relative to $K_{1}$ ($\Delta_{0}^{(3)} = -\Delta_{0}^{(1)}$, $\Delta_{z}^{(3)} = -\Delta_{z}^{(1)}$), reflecting the opposite electric potentials of the outer layers.

The analytical expressions in Eqs.~\eqref{eq:H3G_K1}--\eqref{eq:H3G_K3} capture the essential Dirac physics of h3G in the $\alpha\beta$ domain, and the renormalized Dirac velocities remain in good agreement with the full continuum results for $\theta \gtrsim 2.5^\circ$ [Fig.~\ref{fig:fig2}(c), left].
At smaller twist angles, however, the first-shell parameters deviate systematically from their full continuum counterparts, as the nearest-neighbor truncation is no longer sufficient to accurately reproduce the band structure~\cite{KShin2023}.
In the absence of MDT, the first-shell model yields gapless Dirac points at zero bias, whereas the full continuum model including MDT opens finite intrinsic gaps at $K_{1,2,3}$ [Fig.~\ref{fig:fig2}(c), right].
These zero-bias gaps act as intrinsic Dirac masses that are subsequently tuned by the bias-induced mass terms in Eqs.~\eqref{eq:H3G_K1} and \eqref{eq:H3G_K3}.

\begin{figure}[t!]
\includegraphics[width=1.0\linewidth]{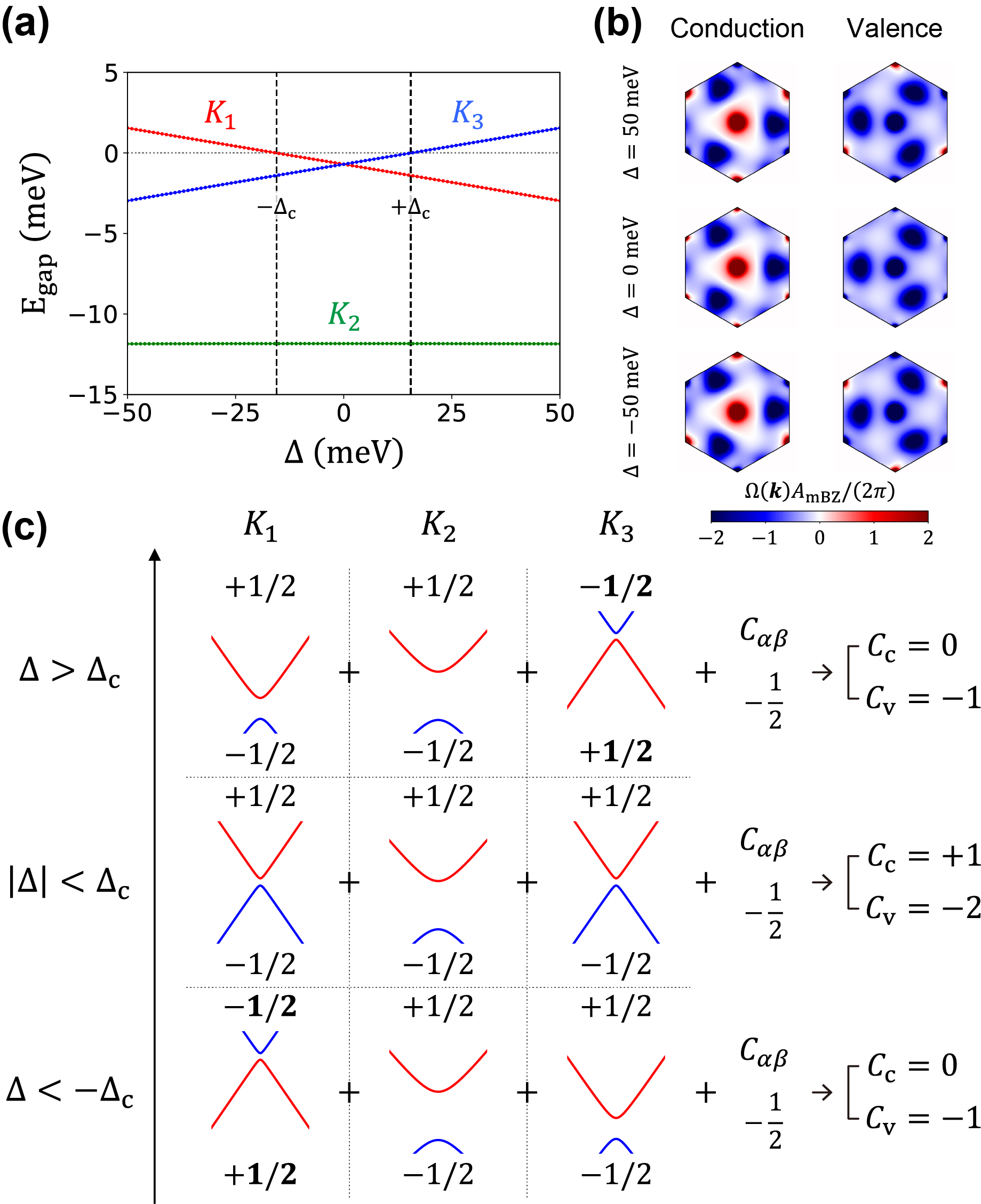}
\caption{
\textbf{Bias-driven evolution of the band topology in h3G.}
(a) Signed Dirac-point gaps at $K_1$ (red), $K_2$ (green), and $K_3$ (blue) as a function of the applied bias $\Delta$ at $\theta=3^{\circ}$.
The $K_1$ and $K_3$ gaps close at $\Delta=\mp\Delta_{\mathrm{c}}$, respectively, defining the critical biases, while the $K_2$ gap remains finite.
(b) Berry-curvature maps of the lowest-energy conduction and valence bands in the mBZ for $\Delta=0$ and $\pm 50$~meV.
Here, $A_{\rm mBZ}$ is the area of the mBZ, and a common color scale is used for all maps.
(c) Schematic illustration of the sector-resolved low-energy topology, with the band colors encoding the sign of the local Berry-curvature contribution.
The numbers above and below denote the local Chern numbers of the conduction and valence bands, respectively, with bold font highlighting the values that change across the topological transitions.
Summing these sector contributions with the constant domain background $C_{\alpha\beta}=-1/2$ yields the net valley Chern numbers $C_{\mathrm{c}}$ and $C_{\mathrm{v}}$.
}
\label{fig:fig3}
\end{figure}

At $\theta=3^\circ$, this competition between MDT and the applied bias produces the signed gap evolution shown in Fig.~\ref{fig:fig3}(a).
At zero bias, the intrinsic gaps are approximately $0.7$~meV at $K_{1,3}$ and $11.8$~meV at $K_{2}$.
The bias-induced mass corrections at the outer-layer sectors $K_{1,3}$ scale as $\alpha^{4}\Delta$ ($\propto \Delta/\theta^{4}$), whereas the middle-layer sector $K_{2}$ has no first-order bias-induced mass under symmetric bias.
Consequently, the small intrinsic masses at $K_{1}$ and $K_{3}$ can be inverted by a moderate electric field, leading to gap closings at opposite bias polarities, while the $K_{2}$ gap remains finite.
The resulting mass signs determine the local Chern contributions of the folded Dirac sectors.

Accordingly, the band Chern number in the $K$ valley can be decomposed into local contributions from the three folded Dirac sectors $K_{1,2,3}$ and a stacking-dependent background associated with the local single-moir\'{e} domain [Fig.~\ref{fig:fig3}(c)].
For $|\Delta|<\Delta_{\rm c}$, the mass terms have the same sign in all three sectors, giving rise to the same sublattice polarization (B/A for conduction/valence bands).
Each sector therefore contributes $+1/2$ ($-1/2$) to the conduction (valence) band.
Combining these contributions with the $\Delta$-independent domain background $C_{\alpha\beta}=-1/2$ gives valley Chern numbers $C_{\mathrm{c}}=+1$ and $C_{\mathrm{v}}=-2$ for the conduction and valence bands, respectively.
When $|\Delta|>\Delta_{\rm c}$, the mass term at either $K_{1}$ or $K_{3}$ flips sign depending on the polarity of $\Delta$, reversing the corresponding local contribution and changing the valley Chern numbers to $C_{\mathrm{c}}=0$ and $C_{\mathrm{v}}=-1$.
This sector-resolved flipping behavior is reflected in the Berry-curvature distributions shown in Fig.~\ref{fig:fig3}(b).

Here, $C_{\alpha\beta}$ denotes the stacking-dependent topological background intrinsic to the local $\alpha\beta$ single-moir\'{e} domain, which underlies the Chern mosaic across the supermoir\'{e} structure~\cite{Guerci2024PRR,Nakatsuji2023,Devakul2023}.
Together with the folded-sector contributions, this background yields integer valley Chern numbers for the isolated moir\'{e} bands.
This sector-resolved Chern-counting picture is supported by lattice tight-binding (TB) calculations using the TAPW method, which reproduce the same Chern-number evolution (see Appendix~\ref{Sec:AppendixA}).
The quantitative difference in $\Delta_{\rm c}$ arises from the residual strain introduced in constructing the real-space supercell, which slightly enhances the zero-bias gap and shifts the critical bias without altering the valley topology.

\subsection{\texorpdfstring{\protect\MakeLowercase{h}4G}{h4G}}
\label{subSec:h4G}

\begin{figure}[t!]
\includegraphics[width=1.0\linewidth]{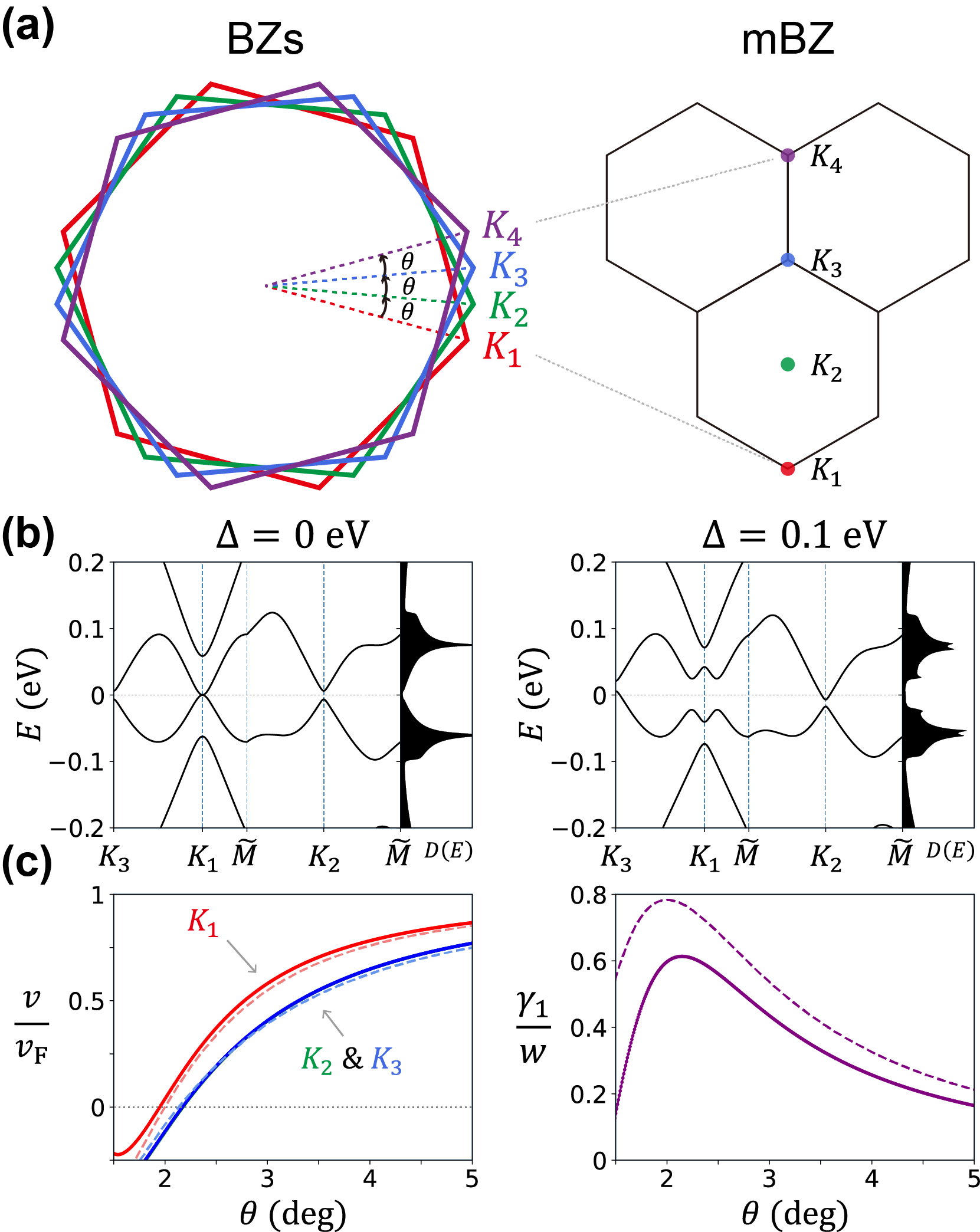}
\caption{
\textbf{Moir\'{e} geometry and low-energy electronic structure of h4G.}
Same as Fig.~\ref{fig:fig2}, but for h4G.
In (a), within the single-moir\'{e} description, the Dirac point $K_{4}$ folds onto $K_{1}$ in the mBZ.
In (c), the right panel shows the twist-angle dependence of the parameter $\gamma_{1}/w$ rather than the MDT-induced gaps.
Red, green, blue, and purple indicate $K_{1}$, $K_{2}$, $K_{3}$, and $K_{4}$, respectively.
}
\label{fig:fig4}
\end{figure}

\noindent
We next turn to h4G, where the additional layer qualitatively changes the folded-sector structure.
In the $\alpha\beta\alpha$ domain, the Dirac point $K_{4}$ folds onto $K_{1}$, hybridizing the outer-layer states into an AB-bilayer-like sector, while the $K_{2}$ and $K_{3}$ sectors remain inner-layer Dirac sectors [Fig.~\ref{fig:fig4}(a,b)].
This mixed structure provides the minimal setting in which monolayer-like and bilayer-like folded sectors coexist within the same reconstructed domain.

The folded $K_{1}$ sector requires an effective four-band description in the outer-layer basis $\{\mathrm{1A},\mathrm{1B},\mathrm{4A},\mathrm{4B}\}$.
Integrating out the inner layers $\mathrm{L}_{2}$ and $\mathrm{L}_{3}$ reduces the first-shell model to
\begin{align}
\label{eq:H4G_K1}
&\mathcal{H}^{K_{1}}_{\alpha\beta\alpha}
=
\begin{pmatrix} 
\Delta^{+}_{z} & v^{(1)}\Pi^{\dagger}_{\bm{k}} & iv_{4}\Pi^{\dagger}_{\bm{k}} & 0 \\
v^{(1)}\Pi_{\bm{k}} & \Delta^{-}_{z} & \gamma_{1} & -iv_{4}\Pi^{\dagger}_{\bm{k}} \\
-iv_{4}\Pi_{\bm{k}} & \gamma_{1} & -\Delta^{-}_{z} & v^{(1)}\Pi^{\dagger}_{\bm{k}} \\
0 & iv_{4}\Pi_{\bm{k}} & v^{(1)}\Pi_{\bm{k}} & -\Delta^{+}_{z}
\end{pmatrix}.
\end{align}
Here, $\Pi_{\bm{k}}=\hbar(k_x+i k_y)$, and the parameters are
\begin{equation}
\label{eq:H4G_K1para}
\begin{aligned}
\frac{\Delta^{\pm}_{z}}{\Delta} &\simeq -\frac{1-4\alpha^{2}+(45\pm 18)\alpha^{4}}{2(1+36\alpha^{4})(1+63\alpha^{4})}, \\
\frac{v^{(1)}}{v_{\rm F}} &= \frac{1-9\alpha^{2}-18\alpha^{4}}
{\sqrt{(1+36\alpha^{4})(1+63\alpha^{4})}}, \\
\frac{\gamma_{1}}{w} &= \frac{18\alpha^{2}(1-3\alpha^{2})}{1+63\alpha^{4}}, \\
\frac{v_{4}}{v_{\rm F}} &= \frac{18\alpha^{3}}{\sqrt{(1+36\alpha^{4})(1+63\alpha^{4})}}.
\end{aligned}
\end{equation}
The coupling $\gamma_{1}$ hybridizes the outer-layer states and produces an AB-bilayer-like parabolic dispersion at $\Delta=0$, while $v_{4}$ provides a subleading correction.
A finite total potential drop $\Delta$ gaps this bilayer-like sector and drives it into a Mexican-hat dispersion [Fig.~\ref{fig:fig4}(b)].

At $K_{2}$ and $K_{3}$, the inner-layer character allows a reduction to effective two-band Dirac models.
For the $\mathrm{L}_{2}$ sector at $K_{2}$, the resulting parameters are
\begin{equation}
\label{eq:H4G_K2}
\begin{aligned}
\frac{v^{(2)}}{v_{\rm F}}
&= \frac{1-12\alpha^{2}}{\sqrt{(1+6\alpha^{2})(1+6\alpha^{2}+27\alpha^{4})}}, \\
\frac{\Delta_{0}^{(2)}}{\Delta} &= -\frac{1+12\alpha^{2}-171\alpha^{4}}{6(1+6\alpha^{2})(1+6\alpha^{2}+27\alpha^{4})}, \\
\frac{\Delta_{z}^{(2)}}{\Delta} &= \frac{18\alpha^{4}}{(1+6\alpha^{2})(1+6\alpha^{2}+27\alpha^{4})}.
\end{aligned}
\end{equation}
The parameters at $K_{3}$ are determined by $C_{2x}$ symmetry.
Unlike h3G, where $C_{2x}$ maps the $\alpha\beta$ domain to $\beta\alpha$, $C_{2x}$ preserves the $\alpha\beta\alpha$ stacking sequence in h4G, exchanging the inner layers ($\mathrm{L}_{2} \leftrightarrow \mathrm{L}_{3}$) and the sublattices ($\mathrm{A} \leftrightarrow \mathrm{B}$).
The symmetry therefore acts within the same domain and gives $v^{(3)} = v^{(2)}$, $\Delta_{0}^{(3)} = -\Delta_{0}^{(2)}$, and $\Delta_{z}^{(3)} = \Delta_{z}^{(2)}$.

Together, these effective models decompose the low-energy structure of h4G into one bilayer-like sector at $K_{1}$ and two monolayer-like inner-layer sectors at $K_{2}$ and $K_{3}$.
The analytical parameters are consistent with those extracted from the full continuum calculations [Fig.~\ref{fig:fig4}(c)]:
the Dirac velocity renormalizations agree quantitatively down to $\theta \approx 2^\circ$, whereas the twist-angle dependence of $\gamma_{1}$ is reproduced at a qualitative level.
Including MDT opens intrinsic gaps at the inner-layer sectors $K_{2}$ and $K_{3}$, while the bilayer-like $K_{1}$ sector remains gapless at zero bias.
This sector hierarchy provides the basis for the gate-tunable topology of h4G.

\begin{figure}[t!]
\includegraphics[width=1.0\linewidth]{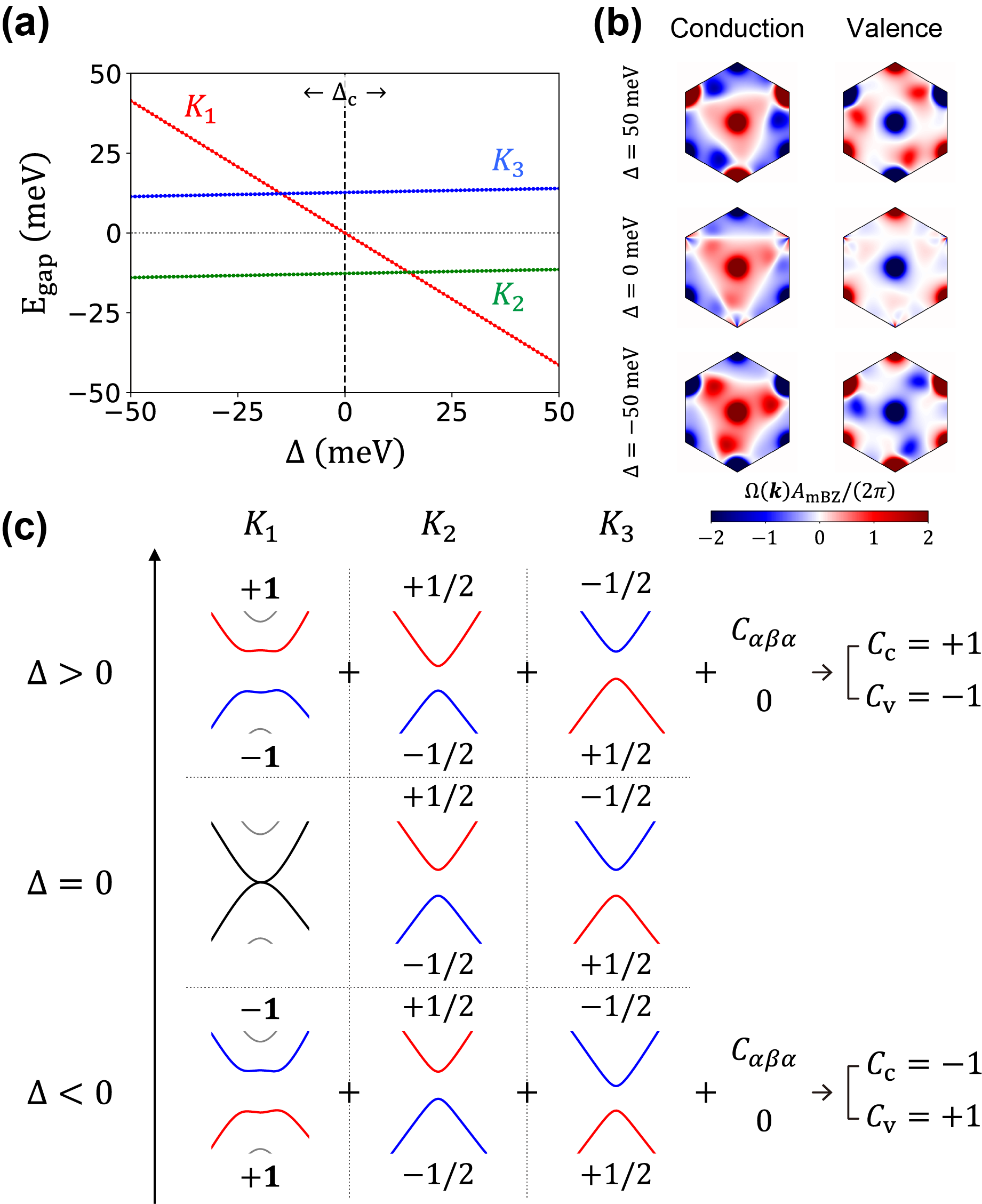}
\caption{
\textbf{Bias-driven evolution of the band topology in h4G.}
Same as Fig.~\ref{fig:fig3}, but for h4G.
Unlike h3G, only the bilayer-like $K_1$ sector undergoes a bias-driven gap closing near $\Delta=0$ in (a), while the inner-layer $K_2$ and $K_3$ gaps remain finite.
Correspondingly, in (c), the bilayer-like $K_1$ sector carries local Chern contributions of $\pm\,\mathrm{sgn}(\Delta)$ with a vanishing domain background ($C_{\alpha\beta\alpha}=0$).
}
\label{fig:fig5}
\end{figure}

At $\theta=3^\circ$, this separation of sectors produces the signed gap evolution shown in Fig.~\ref{fig:fig5}(a).
The intrinsic MDT-induced masses at $K_{2}$ and $K_{3}$ are larger than the weak bias-induced corrections ($\sim \alpha^{4}\Delta$), so the inner-layer sectors remain gapped over the relevant bias range.
By contrast, the bilayer-like $K_{1}$ sector undergoes the bias-driven gap closing near $\Delta=0$, making it the source of gate-tunable valley topology.

The Chern-number decomposition in Fig.~\ref{fig:fig5}(c) reflects this sector hierarchy.
Because $C_{2x}$ acts within the $\alpha\beta\alpha$ domain, the Berry curvature is odd at zero bias, the local contributions from $K_{2}$ and $K_{3}$ cancel, and the domain background vanishes, $C_{\alpha\beta\alpha}=0$.
These cancellations remain intact over the relevant bias range because the inner-layer sectors do not undergo a mass inversion.
The valley Chern number is thus set by the bilayer-like $K_{1}$ sector alone.

To make the $K_{1}$ contribution explicit, we further downfold the four-band Hamiltonian onto the $\{\mathrm{1A},\mathrm{4B}\}$ subspace, treating $v_{4}$ as a higher-order correction.
In this process, we retain only terms linear in $\Delta$ and $v_{4}$, together with the mixed $\Delta v_{4}$ contribution.
This leads to
\begin{equation}
\begin{aligned}
\mathcal{H}_{\rm eff}
&= S\left(h_{\mathrm{ch}} + h_{\mathrm{gap}} + h_{4}\right)S^{\dagger}, \\
h_{\mathrm{ch}}
&= -\frac{(v^{(1)})^{2}}{\gamma_{1}}
\begin{pmatrix}
0 & (\Pi_{\bm{k}}^{\dagger})^{2} \\
\Pi_{\bm{k}}^{2} & 0
\end{pmatrix}, \\
h_{\mathrm{gap}}
&= \left[
\Delta^{+}_{z}
- \frac{\Delta^{-}_{z}(v^{(1)})^{2}}{\gamma_{1}^{2}}
\Pi_{\bm{k}}^{\dagger}\Pi_{\bm{k}}
\right]\sigma_{z}, \\
h_{4}
&= \frac{2\Delta^{-}_{z}v^{(1)}v_{4}}{i\gamma_{1}^{2}}
\begin{pmatrix}
0 & -(\Pi_{\bm{k}}^{\dagger})^{2} \\
\Pi_{\bm{k}}^{2} & 0
\end{pmatrix}, \\
S
&= \left(
1+\frac{(v^{(1)})^{2}}{\gamma_{1}^{2}}
\Pi_{\bm{k}}^{\dagger}\Pi_{\bm{k}}
\right)^{-1/2}\mathbb{I}_{2}.
\end{aligned}
\end{equation}
Here, the leading chiral term $h_{\mathrm{ch}}$ gives the doubled pseudospin chirality and the associated $2\pi$ Berry phase~\cite{McCann2006,Novoselov2006,Min2008a,Min2008b}.
The $v_{4}$ correction preserves this quadratic winding by only rotating the phase of the off-diagonal coupling, while the scalar factor $S$ renormalizes the energy scale without changing the pseudospin structure.
The gate-induced mass $h_{\mathrm{gap}}$ therefore yields valley Chern numbers $C_{\mathrm{c}/\mathrm{v}}=\pm \tau_{z}\,\mathrm{sgn}(\Delta)$ for the lowest-energy conduction and valence bands, respectively, where $\tau_{z}=\pm1$ labels the $K/K'$ valleys.

Real-space lattice calculations reproduce the Chern-number evolution shown in Fig.~\ref{fig:fig5}(c), with the small finite critical bias reflecting the same residual-strain effect discussed for h3G.
This continuum--lattice agreement also highlights the essential role of MDT.
In an MDT-free reference model, the bilayer-like $K_{1}$ sector is still governed by $h_{\mathrm{gap}}$, whereas the inner-layer sectors $K_{2}$ and $K_{3}$ are gapless at $\Delta=0$ and acquire only the bias-induced masses $\Delta_{z}^{(2,3)}$ [Eq.~\eqref{eq:H4G_K2}].
Their same-sign bias evolution would lead, by analytical sign counting, to a $|C|=2$ phase sequence in which the sector contributions add constructively~\cite{Fujimoto2025}.
MDT instead induces opposite zero-bias gaps in the two inner-layer sectors, as constrained by $C_{2x}$, causing their local Chern contributions to cancel.
Thus, MDT is essential for the $|C|=1$ phase sequence of h4G.

\subsection{\texorpdfstring{\protect\MakeLowercase{h}NG}{hNG}}
\label{subSec:hNG}

\begin{figure}[t!]
\includegraphics[width=0.9\linewidth]{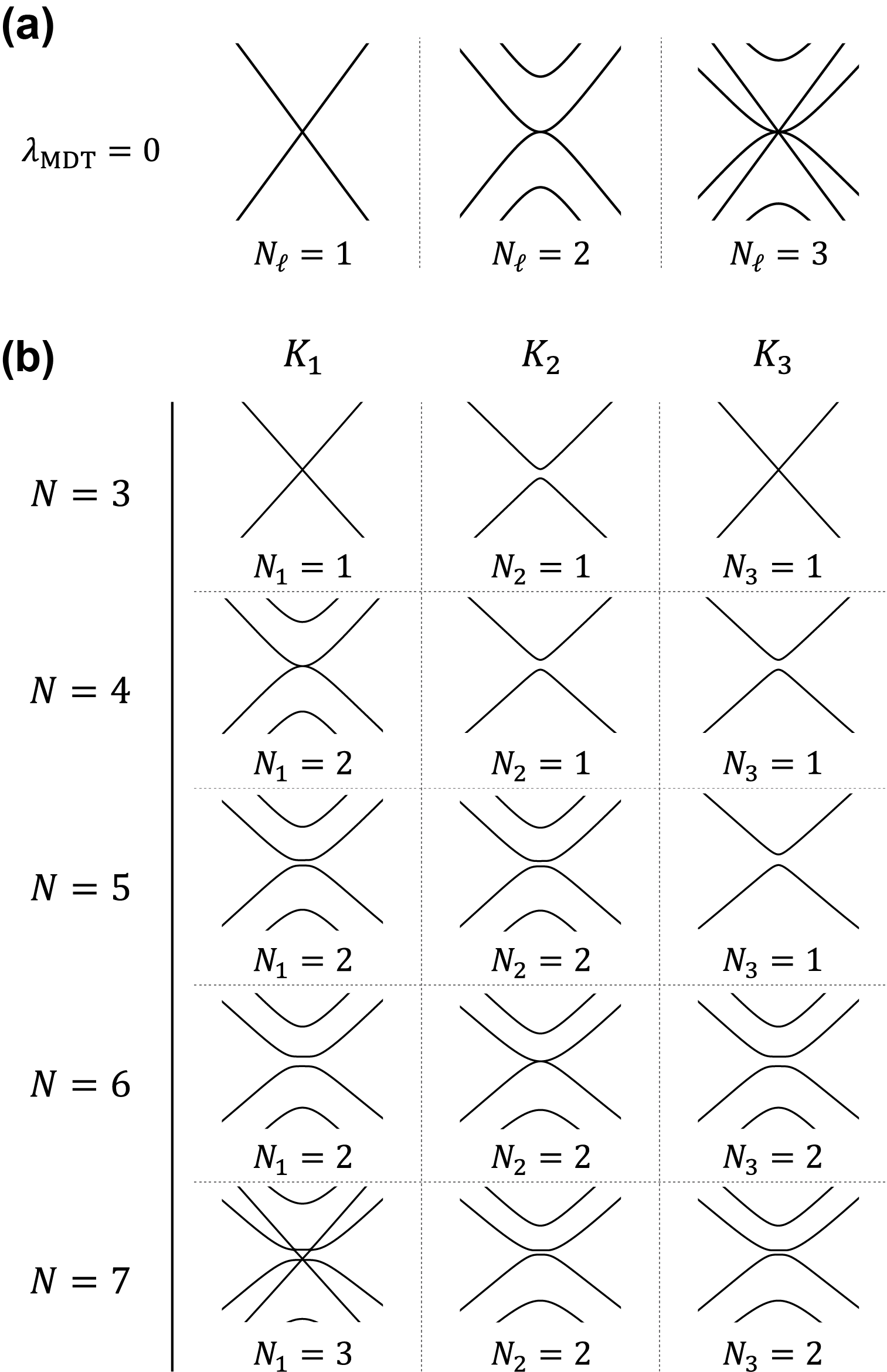}
\caption{
\textbf{Sector-resolved low-energy band structures of h$N$G.}
(a) Schematic low-energy dispersions without the MDT ($\lambda_{\rm MDT} = 0$) for multiplicities $N_{\ell}=1, 2$, and $3$.
(b) Low-energy spectra of the folded sectors $K_{1,2,3}$ for $N=3$--$7$ with MDT included.
Each sector is determined by the number $N_{\ell}$ of layers folded onto the corresponding moir\'{e} momentum and is gapped or reshaped by MDT.
}
\label{fig:fig6}
\end{figure}

\noindent
We finally generalize the folded-sector description to helical $N$-layer graphene within the Bernal-like moir\'{e} stacking family (e.g., $\alpha\beta$, $\alpha\beta\alpha$, \ldots).
The $C_{2z}T$-related counterpart (e.g., $\beta\alpha$, $\beta\alpha\beta$, \ldots) follows from the same construction.
Within the single-moir\'{e} approximation, the layer Dirac points fold onto the three mBZ points $K_{\ell}$, with $\ell=1,2,3$ labeling the folded sectors.
These sectors collect layers $(\mathrm{L}_{1},\mathrm{L}_{4},\mathrm{L}_{7},\ldots)$, $(\mathrm{L}_{2},\mathrm{L}_{5},\mathrm{L}_{8},\ldots)$, and $(\mathrm{L}_{3},\mathrm{L}_{6},\mathrm{L}_{9},\ldots)$, respectively.
Consequently, the low-energy theory decomposes into three folded sectors, each described by an effective $2N_{\ell}\times2N_{\ell}$ Hamiltonian with the structure of Bernal-stacked graphene, where $N_{\ell}$ is the number of layers folded onto $K_{\ell}$.
Figure~\ref{fig:fig6} summarizes the resulting band character for $N=3\text{--}7$, with and without MDT.

Without MDT, the sector decomposition maps the $K_{1}$ and $K_{3}$ sectors onto an $\mathrm{ABA}\cdots$ Bernal sequence and the $K_{2}$ sector onto a $\mathrm{BAB}\cdots$ sequence.
Including MDT and a perpendicular electric field opens gaps at the low-energy Dirac points and induces sublattice polarization in the corresponding states.
For the monolayer-like sectors ($N_{\ell}=1$), the local Chern contributions are
$C_{K_{\ell}, \, \mathrm{c}/\mathrm{v}}=\pm\frac{1}{2}\tau_{z}\,\mathrm{sgn}(\Delta-\Delta_{\mathrm{c},\ell})$,
where $\Delta_{\mathrm{c},\ell}$ denotes the critical bias at which the $K_{\ell}$-sector gap closes.
Because this gap depends weakly on $\Delta$, bias-driven gap closing can occur only in sectors with small MDT masses, such as the $K_{1}$ and $K_{3}$ sectors of h3G.
In contrast, the bilayer-like sectors ($N_{\ell}=2$) exhibit gaps that are strongly tuned by the interlayer bias.
Sweeping the bias through the critical point $\Delta_{\mathrm{c},\ell}$ reverses the sign of the effective mass, closing and reopening the gap and thereby driving a topological transition.
Here, MDT shifts $\Delta_{\mathrm{c},\ell}$ slightly away from zero without qualitatively changing this bias-driven evolution.
The local Chern contributions are
$C_{K_{\ell}, \, \mathrm{c}/\mathrm{v}}=\pm(-1)^{\ell+1}\tau_{z}\,\mathrm{sgn}(\Delta-\Delta_{\mathrm{c},\ell})$,
where the factor $(-1)^{\ell+1}$ reflects the opposite stacking sequence of the $K_{2}$ sector.

For the Bernal-like stacking family, the domain contribution is fixed by layer parity, vanishing for even $N$ and taking the value $-1/2$ for odd $N$, consistent with the action of $C_{2x}$ symmetry on the local single-moir\'{e} domain.
This background contribution combines with the sector-resolved local contributions to obtain the net valley Chern number of the isolated moir\'{e} bands.
Together with the layer-folding rules summarized in Fig.~\ref{fig:fig6}, these results show how the $\alpha\beta\alpha$ family in helical multilayers hosts gate-tunable topological bands.

\section{Discussion}
\label{Sec:Sec4}

\noindent
Our study establishes a domain-resolved picture of the low-energy electronic structure and valley topology of helical multilayer graphene.
Observing that lattice relaxation partitions the supermoir\'{e} pattern into distinct domains, we have identified moir\'{e}-scale stacking orders in h4G: AA-like ($\alpha\alpha\alpha$), Bernal-like ($\alpha\beta\alpha$), and rhombohedral-like ($\alpha\beta\gamma$).
These local domain types naturally generalize to h$N$G, forming the corresponding stacking families.
Crucially, we have shown using the analytical first-shell model that the low-energy spectrum decomposes into three folded Dirac sectors, governed by the Dirac-cone multiplicities $N_{\ell}$ (e.g., $2+1+1$ decomposition for h4G and $2+2+1$ for h5G).

Once the low-energy spectrum is partitioned by $N_{\ell}$, the band character is determined by the local moir\'{e} displacement vectors $\bm{d}_{\ell\ell'}$, which specify the stacking sequence.
As shown above, the $\alpha\beta\alpha$ family reduces to Bernal-stacked $N_{\ell}$-layer graphene in each sector.
Similarly, the $\alpha\alpha\alpha$ family connects to AA-stacked multilayer graphene, whereas the $\alpha\beta\gamma$ family yields a low-energy band structure that cannot be reduced to rhombohedral multilayer graphene, as detailed in Appendix~\ref{Sec:AppendixB}.

\begin{figure}[t!]
\includegraphics[width=1.0\linewidth]{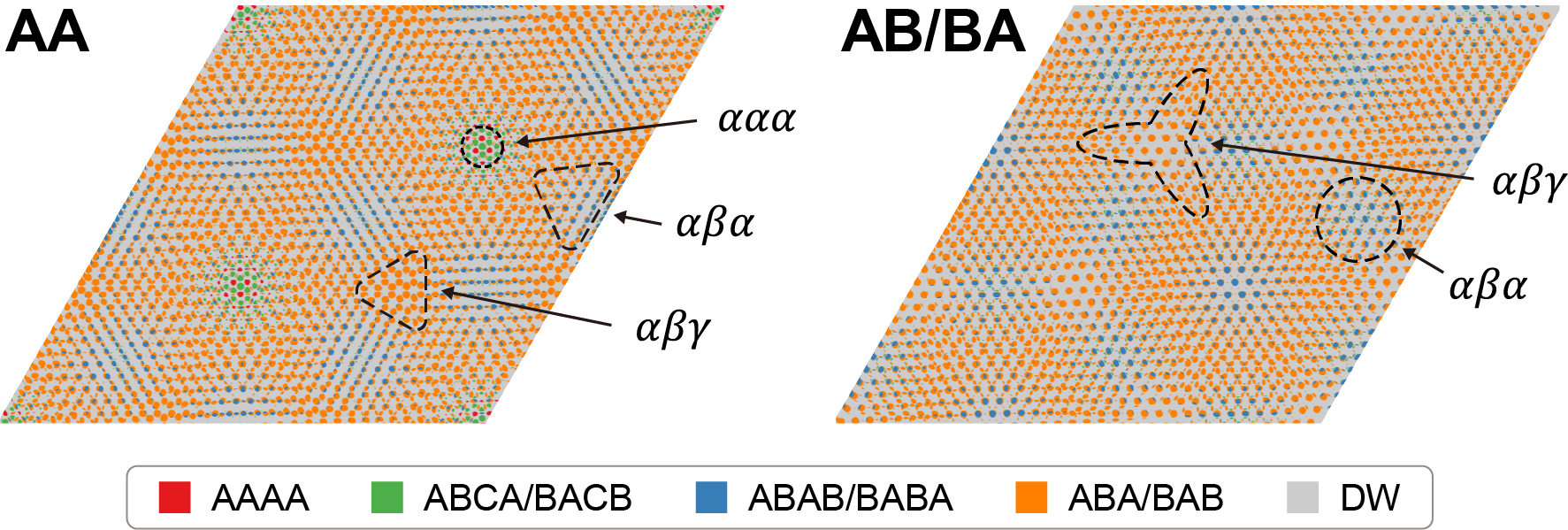}
\caption{
\textbf{Domain structures in h4G with top-layer sliding.}
Relaxed supermoir\'e-scale stacking maps at $\theta=3^{\circ}$ for the unslid AA (left) and slid AB/BA (right) initial top-layer configurations.
The two relaxed textures contain the same stacking families, indicating that the supermoir\'{e} tessellation is largely insensitive to the initial top-layer sliding.
The main residual difference is a minute $\alpha\alpha\alpha$ domain near the supermoir\'{e} corner in the unslid AA case.
}
\label{fig:fig7}
\end{figure}

To assess this domain-resolved picture beyond the continuum and effective models, we perform lattice TB calculations and apply the TAPW method to obtain valley-resolved bands and evaluate their Chern numbers.
Overall, the lattice calculations reproduce the characteristic domain-dependent band structures and valley topology obtained from the continuum and effective models (Figs.~\ref{fig:figA1} and \ref{fig:figA2}).
The MDT term in the continuum Hamiltonian is essential for defining the topological character and bringing the continuum and lattice results into closer agreement.
Specifically, while the Dirac nodes in the $\alpha\alpha\alpha$ family remain gapless, the bands in the $\alpha\beta\alpha$ and $\alpha\beta\gamma$ families become gapped and acquire a pronounced sublattice character.
Accordingly, the net valley Chern number is determined by the combined contributions of the individual sectors and the domain background.

In principle, increasing the layer number $N$ requires progressively larger supercells to accommodate successive moir\'{e} patterns.
Here, however, we restrict our analysis to the supermoir\'{e} scale and disregard longer-range interference effects that may appear for $N>3$.
To test this approximation, we rigidly shift the top layer to generate AB and BA initial configurations (Fig.~\ref{fig:fig7}), thereby sampling representative local registries of the neglected longer-range interference pattern.
Upon relaxation, the same stacking families appear in both the unslid and slid cases.
The remaining difference is a minute, topologically trivial $\alpha\alpha\alpha$ domain at the supermoir\'{e} corner in the unslid configuration, whereas the $\alpha\beta\alpha$ and $\beta\alpha\beta$ stacking families differ only slightly in their spatial connectivity.
The near-indistinguishability of the resulting domain patterns shows that longer-range variations beyond the supermoir\'{e} scale likely do not generate additional nontrivial domains.
Therefore, we posit that the supermoir\'{e} lattice provides a valid basis for the domain-resolved continuum description.

\begin{figure}[t!]
\includegraphics[width=1.0\linewidth]{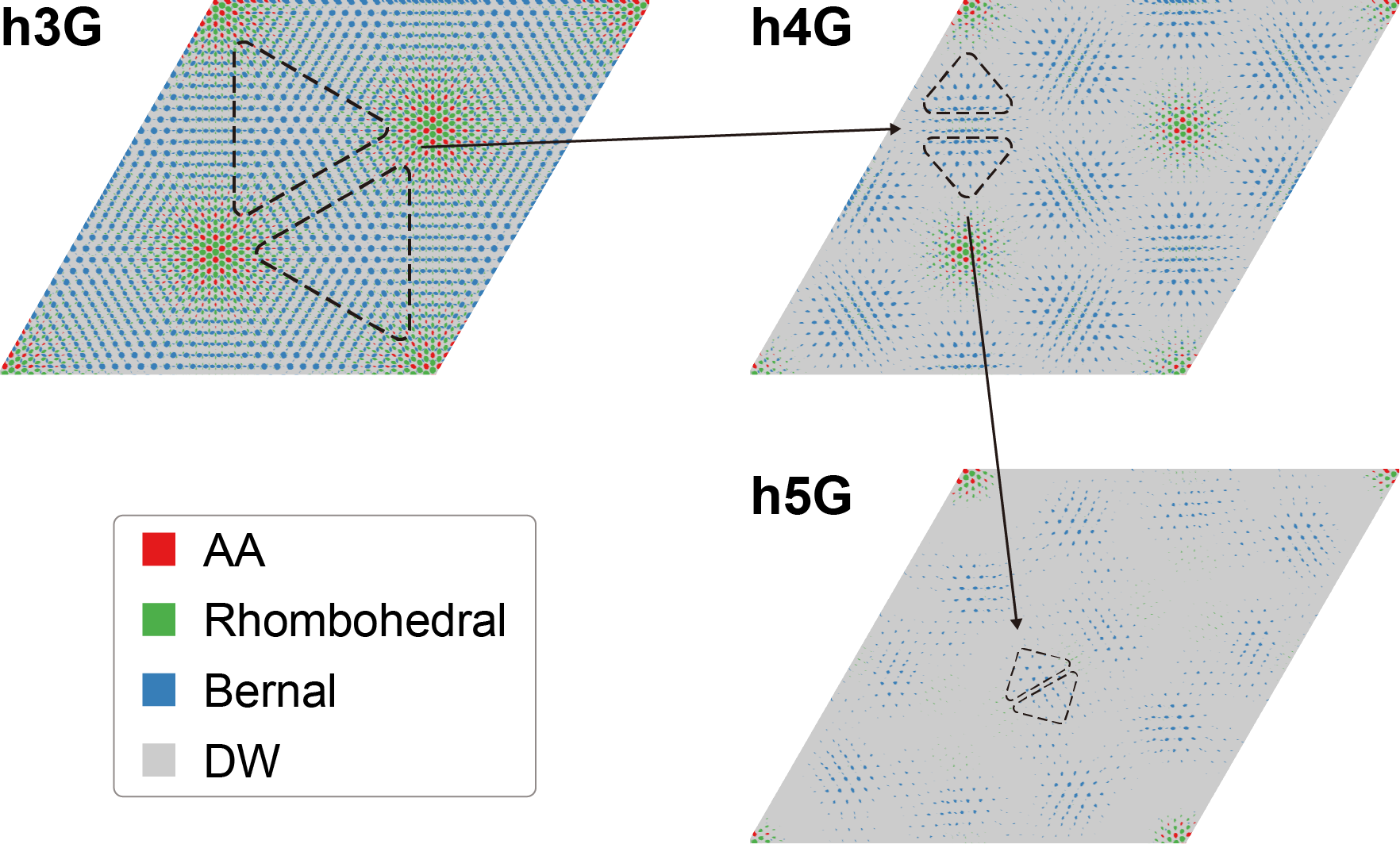}
\caption{
\textbf{Progressive fragmentation of single-moir\'{e} domains with layer number.}
Relaxed stacking maps for h3G (top left), h4G (top right), and h5G (bottom right) at $\theta=3^{\circ}$.
As the layer number $N$ increases, the stacking domains become progressively fragmented and the structure becomes increasingly dominated by domain-wall (DW) regions.
Dashed outlines and solid arrows track the hierarchical size reduction of the $\alpha\beta\alpha$ and $\beta\alpha\beta$ families across the three systems.
For clarity, the additional stacking class shown in orange in Figs.~\ref{fig:fig1} and \ref{fig:fig7} is not shown separately here.
}
\label{fig:fig8}
\end{figure}

The structural complexity intensifies in thicker stacks.
In h5G, the additional moir\'{e} interface leads to a domain tessellation that is far more intricate than in h3G or h4G.
As shown in Fig.~\ref{fig:fig8}, well-resolved local stacking regions progressively lose their spatial coherence, while the area fraction of DW regions increases markedly.
Beyond h4G, where the cumulative rotation between outermost layers reaches $9^\circ$, only sparse and spatially confined stacking regions remain, corresponding to single-moir\'{e} stacking motifs.
This domain fragmentation establishes a practical bound for our model, suggesting that the single-moir\'{e} approximation becomes unreliable once $(N-1)\theta \gtrsim 10^\circ$.

Finally, the recent experimental realization of topological bands and correlated states in h3G~\cite{LXia2025,Bocarsly2026,Hoke2026} highlights the relevance of h$N$G systems as a platform for moir\'{e} quantum matter.
Notably, our results show that local moir\'{e} stacking sequences in h$N$G can generate emergent multilayer-like sectors, even though the underlying system is not necessarily conventionally stacked multilayer graphene.
This sector-resolved picture further implies distinct bias fingerprints across h$N$G domains, such as gate-tunable gaps in the $\alpha\beta\alpha$ stacking family.
In this context, our domain-resolved continuum framework, along with the analytical decomposition rules for each stacking family, provides a natural starting point for exploring thicker helical stacks.

\section*{Acknowledgements}
\noindent
This work was supported by the National Research Foundation of Korea (NRF) through Grants RS-2023-NR076715 (K.S. and H.M.), RS-2023-00249414 (N.L.), and NRF-2020R1A5A1016518 (J.J.).
K.S. and H.M. acknowledge support from the Creative-Pioneering Researchers Program at SNU and the Center for Theoretical Physics.
N.L. and J.J. acknowledge computational support from KISTI Grant KSC-2022-CRE-0514 and the use of the resources of Urban Big Data and AI Institute (UBAI) at UOS.

\section*{Data availability statement}
\noindent
The data that support the findings of this study are available upon reasonable request from the authors.

\appendix

\renewcommand{\thefigure}{A\arabic{figure}}
\renewcommand{\theHfigure}{A\arabic{figure}}
\setcounter{figure}{0}

\renewcommand{\thesection}{\Alph{section}}
\setcounter{section}{0}
\makeatletter
\renewcommand{\p@subsection}{}
\makeatother
\titleformat{\section}[block]{\normalfont\large\bfseries}{Appendix~\Alph{section}.}{0.5em}{}

\renewcommand{\thesubsection}{\thesection.\arabic{subsection}}
\titleformat{\subsection}[block]{\normalfont\normalsize\bfseries}{\thesubsection.}{0.5em}{}

\section{Lattice calculations}
\label{Sec:AppendixA}

\noindent
In this Appendix, we provide details of the real-space lattice calculations used to obtain the band structure and valley Chern numbers of helical multilayer graphene.
We also detail the extraction of stacking maps based on the local interlayer distances, which characterize the local stacking configuration.
While the continuum model in the main text captures the low-energy electronic structure, the real-space lattice calculations reported here provide an independent benchmark.
They retain atomistic details, including strain-modified intralayer hoppings and relaxation-induced variations of the local stacking and interlayer tunneling.

\subsection{Tight-binding model and band structures}
\label{subSec:A1}

\begin{figure*}[tbp]
\includegraphics[width=0.75\linewidth]{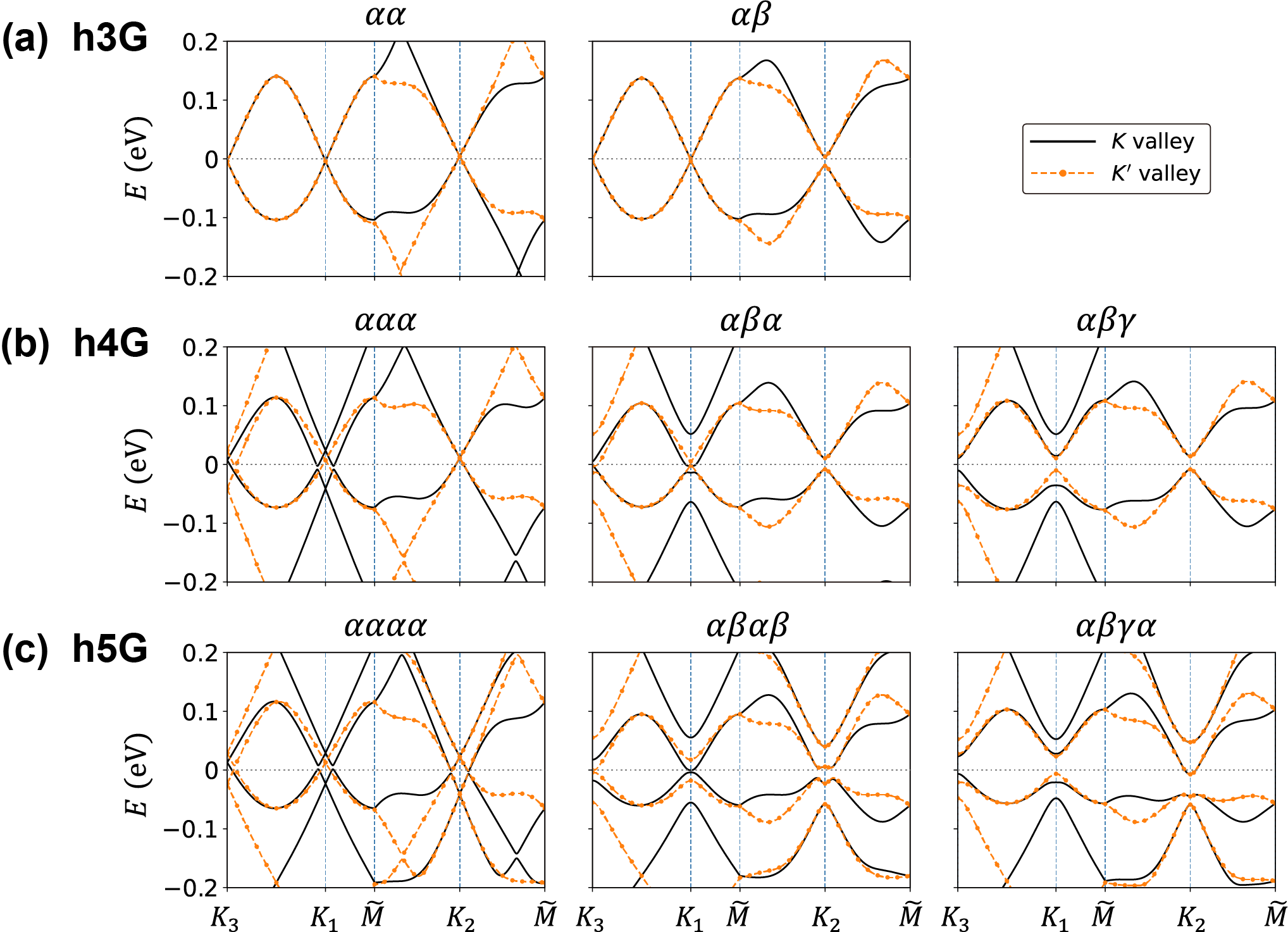}
\caption{
Representative lattice TB band structures for the $\alpha\alpha\alpha$, $\alpha\beta\alpha$, and $\alpha\beta\gamma$ stacking families in (a) h3G, (b) h4G, and (c) h5G.
In the real-space lattice construction, both valleys coexist at the folded momenta, and the valley-resolved bands (solid black for $K$ and dashed orange for $K'$) are extracted using the TAPW method described in Supplementary Sec.~\ref{subSec:A3}.
Overall, the lattice spectra show good qualitative agreement with the continuum results in the main text, with small quantitative differences in the Dirac-point energies and gap sizes.
}
\label{fig:figA1}
\end{figure*}

\noindent
To complement the continuum band structures presented in the main text, we show in Fig.~\ref{fig:figA1} representative lattice band structures for the $\alpha\alpha\alpha$, $\alpha\beta\alpha$, and $\alpha\beta\gamma$ stacking families in 3-, 4-, and 5-layer systems, computed within the scaled hybrid exponential (SHE) tight-binding (TB) model~\cite{Leconte2022TBG}.
In this model, the intralayer and interlayer hoppings are treated separately.
The intralayer Hamiltonian is described by the F2G2 parametrization~\cite{Jung2013}, which retains the structure factors $f_{n}$ and $g_{n}$ up to $n=2$ for the graphene $\pi$ bands.
The parameters are obtained from Wannierized LDA DFT, reproducing the $\pi$-band structures and yielding an effective Fermi velocity of $v_{\rm F} \simeq 10^6~\mathrm{m/s}$.
To account for lattice relaxation, strain effects are incorporated via an exponential rescaling of the intralayer hopping amplitudes~\cite{Pereira2009},
\begin{equation}
t^{\mathrm{intra}}_{ij} = t^{\mathrm{F2G2}}_{ij} 
\exp\!\left[-\beta \left(\frac{r_{ij}-r_{ij}^0}{r_{ij}^0}\right)\right],
\end{equation}
where $r_{ij}$ and $r_{ij}^0$ denote the relaxed and reference bond lengths, respectively, and $\beta=3.37$ is the decay parameter.

The interlayer hopping is modeled by a two-center Slater--Koster form with $\pi$ and $\sigma$ components, $t^{\mathrm{TC}}_{ij}$, which is further modified by an exponential dependence on the local interlayer distance,
\begin{equation}
t^{\mathrm{inter}}_{ij} 
= S \exp\!\left(\frac{c_{ij}-p}{q}\right) 
t^{\mathrm{TC}}_{ij},
\end{equation}
where $c_{ij}$ is the vertical separation between sites, $p$ and $q$ are fitted parameters, and $S=0.895$ is a relaxation-dependent scaling factor chosen to calibrate the effective magic angle in t2G.
This hybrid formulation allows for an accurate description of both strain-induced intralayer renormalization and stacking-dependent interlayer tunneling.

To construct a single-moir\'{e} lattice description in real space, we impose a small uniform in-plane strain on each layer such that the layer-resolved Dirac points $K_{\ell}$ become mutually aligned in reciprocal space.
The magnitude of the required strain decreases with decreasing twist angle, so this construction becomes increasingly accurate in the small-angle regime and for a limited number of layers.

Despite the strain introduced in the lattice construction, the resulting band structures remain in strong qualitative agreement with the continuum results.
The main differences are small but systematic, including slight offsets of the Dirac-point energies and finite gaps at the $K_{\ell}$ points already at zero bias, as well as minor quantitative deviations in the extracted band gaps and the critical bias $\Delta_{\mathrm{c}}$.
These discrepancies can be traced back to the residual strain-induced lattice mismatch between layers in the real-space supercell, together with the slight deviations of the $\mathrm{L}_{4}$ and $\mathrm{L}_{5}$ twist angles from the nominal $3^\circ$ value (Table~\ref{tab:MasterSupercell}).
Importantly, these effects do not modify the valley Chern numbers, so the domain-resolved continuum model remains sufficient to capture the relevant real-space physics.

\begin{table*}[t]
\centering
\caption{Integer pairs $(n_{\ell},m_{\ell})$ and corresponding geometric parameters for the single-moir\'{e} and supermoir\'{e} constructions.
Layers are ordered by increasing absolute twist angle $\theta_{\ell}$.
We report the lattice constant $a_{\ell}$, the moir\'{e} or supermoir\'{e} period $L$,
the lattice mismatch with respect to the $0^{\circ}$ reference layer
$\delta_{1\ell}=\left(a_{\ell}-a_{1}\right)/a_{1}\times 100\%$,
the absolute twist angle $\theta_{\ell}$, the incremental twist
$\Delta\theta_{\ell}=\theta_{\ell}-\theta_{\ell-1}$, and the number of atoms per layer $N^{(\ell)}_{\mathrm{at}}$.}
\label{tab:MasterSupercell}
\begin{ruledtabular}
\begin{tabular}{l l c c c c c c c c}
System 
& Layer 
& $(n_{\ell},m_{\ell})$
& $a_{\ell}$ (\AA)
& $L$ (nm)
& $\delta_{1\ell}$ (\%)
& $\theta_{\ell}$ (deg)
& $\Delta\theta_{\ell}$ (deg)
& $N^{(\ell)}_{\mathrm{at}}$ \\ \hline

\multicolumn{9}{l}{\textbf{Single-moir\'{e}}} \\

       & $\mathrm{L}_{1}$ & $(10,12)$ & $2.460190$ & $4.6937$ & $+0.000000$ & $0.000000$ & -- & $728$ \\
       & $\mathrm{L}_{2}$ & $(11,11)$ & $2.463576$ & $4.6937$ & $+0.137632$ & $3.004492$ & $3.004492$ & $726$ \\
       & $\mathrm{L}_{3}$ & $(12,10)$ & $2.460190$ & $4.6937$ & $+0.000000$ & $6.008983$ & $3.004491$ & $728$ \\
       & $\mathrm{L}_{4}$ & $(13,9)$  & $2.450114$ & $4.6937$ & $-0.409562$ & $8.997042$ & $2.988059$ & $734$ \\
       & $\mathrm{L}_{5}$ & $(14,8)$  & $2.433593$ & $4.6937$ & $-1.081095$ & $11.952767$ & $2.955725$ & $744$ \\[6pt]

\multicolumn{9}{l}{\textbf{Supermoir\'{e}}} \\

       & $\mathrm{L}_{1}$ & $(335,402)$ & $2.460190$ & $157.2404$ & $+0.000000$ & $0.000000$ & -- & $816998$ \\
       & $\mathrm{L}_{2}$ & $(369,369)$ & $2.460238$ & $157.2404$ & $+0.001951$ & $3.004492$ & $3.004492$ & $816966$ \\
       & $\mathrm{L}_{3}$ & $(402,335)$ & $2.460190$ & $157.2404$ & $+0.000000$ & $6.008983$ & $3.004491$ & $816998$ \\
       & $\mathrm{L}_{4}$ & $(434,300)$ & $2.460018$ & $157.2404$ & $-0.006991$ & $9.021355$ & $3.012372$ & $817112$ \\
       & $\mathrm{L}_{5}$ & $(464,265)$ & $2.460244$ & $157.2404$ & $+0.002195$ & $11.960818$ & $2.939463$ & $816962$ \\

\end{tabular}
\end{ruledtabular}
\end{table*}

\subsection{TAPW method and valley Chern numbers}
\label{subSec:A2}

\noindent
To compute valley-resolved Chern numbers, we first construct valley-resolved electronic bands using the truncated atomic plane-wave (TAPW) method~\cite{Miao2023}.
In this approach, the real-space TB Hamiltonian is projected onto a momentum-space basis composed of atomic plane waves centered around a chosen valley, $K$ or $K^{\prime}$.
The basis states take the form
\begin{equation}
\bigl|\psi_n^\sigma(\bm{k})\bigr\rangle
=
\frac{1}{\sqrt{N_\sigma}}
\sum_{i}
e^{ i(\bm{k}+\bm{G}_n)\cdot\bm{r}_i^\sigma }
\bigl|\varphi_i^\sigma\bigr\rangle ,
\label{eq:PW}
\end{equation}
where $\bigl | \varphi_{i}^{\sigma} \bigr\rangle$ denotes a localized atomic orbital on sublattice $\sigma$ in the $i$-th atomic unit cell, $\bm{r}_{i}^\sigma$ is its real-space position, and $N_\sigma$ is the number of atoms belonging to sublattice $\sigma$ within the supermoir\'{e} unit cell.
The vectors $\bm{G}_{n}$ are reciprocal lattice vectors of the supermoir\'{e} lattice, and the TAPW truncation retains only those basis functions for which $\bm{k}+\bm{G}_{n}$ lies within a cutoff region surrounding the $K$ or $K^{\prime}$ valley.
Here, the cutoff includes the first 5 shells of $\bm{G}_n$ vectors.

For each crystal momentum $\bm{k}$ in the supermoir\'{e} Brillouin zone, the projected Hamiltonian is constructed as
\begin{equation}
H_{\mathrm{TAPW}}(\bm{k}) = X^\dagger(\bm{k})\, H_{\mathrm{TB}}\, X(\bm{k}),
\end{equation}
where $H_{\mathrm{TB}}$ is the real-space TB Hamiltonian and $X(\bm{k})$ is the projection matrix with elements
\begin{equation}
\bigl(X_\sigma(\bm{k})\bigr)_{i,n}
=
\frac{1}{\sqrt{N_\sigma}}
e^{ i(\bm{k}+\bm{G}_n)\cdot\bm{r}_i^\sigma } .
\end{equation}
In the absence of truncation, this transformation is unitary; in practice, the truncation near a chosen valley yields an efficient low-energy representation of the Hamiltonian.
Diagonalizing the TAPW Hamiltonian produces the eigenvalues $E_{\lambda}(\bm{k})$ and eigenstates $| w_{\lambda\bm{k}}\rangle$.

These momentum-space eigenstates encode the band topology, which we quantify by evaluating the Berry curvature via the Kubo formula~\cite{Wang2006}:
\begin{equation}
\Omega_{\lambda}(\bm{k})
=
-2\hbar^{2}\,\mathrm{Im}
\sum_{\lambda'\neq \lambda}
\frac{
\langle w_{\lambda\bm{k}} | \hat{v}_x(\bm{k}) | w_{\lambda'\bm{k}} \rangle
\langle w_{\lambda'\bm{k}} | \hat{v}_y(\bm{k}) | w_{\lambda\bm{k}} \rangle
}{
\bigl[E_{\lambda'}(\bm{k})-E_\lambda(\bm{k})\bigr]^2
},
\end{equation}
where $\hat{v}_\mu(\bm{k}) = \hbar^{-1}\,\partial H_{\mathrm{TAPW}}(\bm{k})/\partial k_{\mu}$ defines the velocity operators for $\mu = x, y$.
The valley Chern number associated with band $\lambda$ is then obtained by integrating the Berry curvature over the supermoir\'{e} Brillouin zone,
\begin{equation}
C_{\lambda}
=
\frac{1}{2\pi}
\frac{A_{\mathrm{BZ}}}{N_k}
\sum_{\bm{k}\in \mathrm{BZ}}
\Omega_{\lambda}(\bm{k}),
\label{eq:Chern}
\end{equation}
where $A_{\mathrm{BZ}}$ is the Brillouin-zone area and $N_{k}$ is the total number of sampled $\bm{k}$ points.
We perform this discrete momentum-space integration on a uniform $144 \times 144$ Monkhorst--Pack grid, which ensures convergence of the topological invariants.

\begin{figure}[!tbp]
\includegraphics[width=1.0\linewidth]{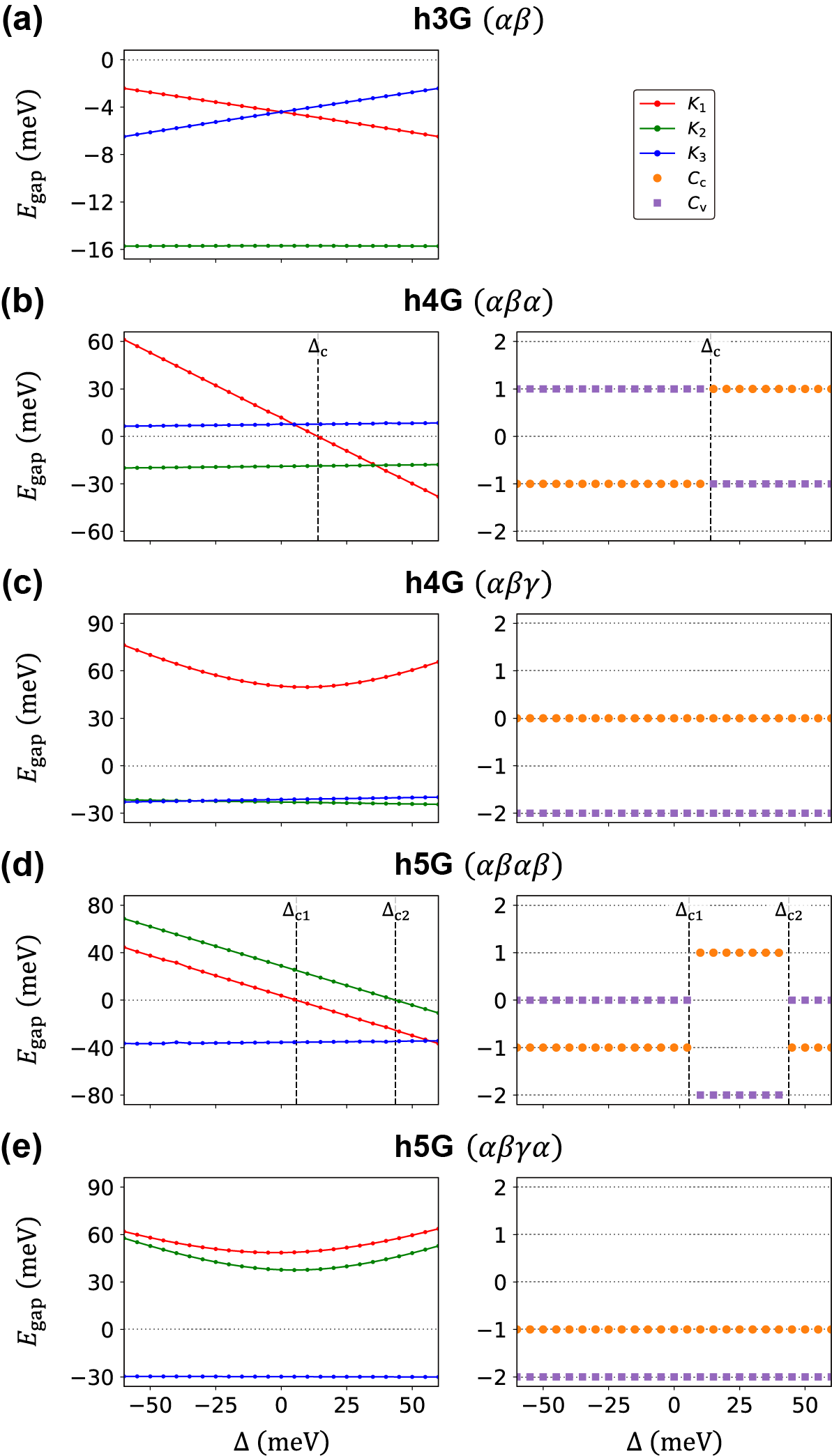}
\caption{
    Bias-driven evolution of the Dirac-point gaps and valley Chern numbers from real-space lattice calculations for (a) h3G ($\alpha\beta$), (b) h4G ($\alpha\beta\alpha$), (c) h4G ($\alpha\beta\gamma$), (d) h5G ($\alpha\beta\alpha\beta$), and (e) h5G ($\alpha\beta\gamma\alpha$).
    The left panels show the signed gaps at $K_1$ (red), $K_2$ (green), and $K_3$ (blue) as functions of the applied bias $\Delta$, where the sign tracks band inversions following the continuum-model convention.
    The right panels show the corresponding valley Chern numbers of the lowest-energy conduction ($C_{\mathrm{c}}$, purple circles) and valence ($C_{\mathrm{v}}$, orange squares) bands.
    For h3G, the valley Chern numbers are omitted because the extremely small gaps make their numerical evaluation unreliable.
    Vertical dashed lines indicate the critical fields, whose shifts relative to the continuum predictions arise mainly from single-moir\'{e} strain.
}
\label{fig:figA2}
\end{figure}

Figure~\ref{fig:figA2} illustrates the bias evolution of the signed Dirac-point gaps and, where numerically reliable, the corresponding valley Chern numbers from the real-space lattice calculations.
The results are in close agreement with the continuum predictions.
Quantitatively, the single-moir\'{e} strain mainly shifts the critical field $\Delta_{\mathrm{c}}$, while the specific choice of TB model and energy minimization force-field primarily affects the magnitude of the gaps.
We do not display the h3G valley Chern numbers, because the extremely small band gaps make numerical convergence challenging, although the same trend is reproduced.
The corresponding critical field, $\Delta_{\mathrm{c}} \approx 130$~meV, also lies outside the bias range of Fig.~\ref{fig:figA2}.

\subsection{Stacking-map construction and classification}
\label{subSec:A3}

\noindent
To characterize the complex spatial textures of the relaxed multilayer structures, we construct stacking maps by assigning each atomic position to a discrete high-symmetry stacking class based on its local interlayer registry, as defined in Ref.~\cite{Li2024}.
For a given pair of adjacent layers $(\ell, \ell+1)$, we separate the atoms into A and B sublattices and compute the local interlayer distances as
\begin{equation}
d^{(\ell)}_{\sigma\sigma'}(\bm{r}) =
\min_{\bm{r}' \in \sigma'_{\ell+1}}
\left| \bm{r} - \bm{r}' \right|,
\qquad \sigma,\sigma' \in \{\mathrm{A},\mathrm{B}\},
\end{equation}
where $\bm{r}$ labels an atom on sublattice $\sigma$ of layer $\mathrm{L}_{\ell}$, and the minimum is taken over all atoms on sublattice $\sigma'$ of layer $\mathrm{L}_{\ell+1}$.
We interpolate these distances onto a dense spatial grid encompassing the moir\'{e} unit cell and assign local stacking labels using the geometric threshold
\begin{equation}
d_0 = \sqrt{\frac{\sqrt{3}}{8\pi}}\, a.
\end{equation}
A point $\bm{r}$ is classified as AA stacking if $d^{(\ell)}_{\mathrm{AA}}(\bm{r}) < d_{0}$ or $d^{(\ell)}_{\mathrm{BB}}(\bm{r}) < d_{0}$, and as AB (BA) stacking if $d^{(\ell)}_{\mathrm{BA}}(\bm{r}) < d_{0}$ ($d^{(\ell)}_{\mathrm{AB}}(\bm{r}) < d_{0}$). 
Points that satisfy none of these criteria are classified as domain walls (DWs).
This choice of $d_{0}$ ensures that, in a rigid twisted homobilayer, the associated AA, AB, BA, and DW regions occupy equal areas.
Although relaxed structures naturally deviate from ideal high-symmetry registries, this thresholding robustly identifies the locally dominant stacking environment.

For multilayer systems, the stacking labels of successive interfaces $(\ell,\ell+1)$ are combined into ordered tuples $(l_{12}, l_{23}, \ldots)$, which define the higher-order stacking motifs used to label our stacking maps.
The DW label is assigned whenever any of the constituent stacking conditions is not satisfied.
This coarse-grained classification does not imply exact local commensurability, but rather provides a physically transparent discretization of the continuously distorted lattice.
By mapping relaxed atomic configurations onto discrete stacking classes, this approach enables a quantitative comparison between rigid and relaxed structures and facilitates a systematic analysis of moir\'{e} and supermoir\'{e} patterns across multiple length scales.

With the validity of the supermoir\'{e} description established, the deviations between the idealized single-moir\'{e} representations [Fig.~\ref{fig:fig1}(b,c)] and the local stacking maps within the supermoir\'{e} lattice [Fig.~\ref{fig:fig1}(a)] can be understood as a supermoir\'{e} modulation of the underlying patterns.
In the single-moir\'{e} construction, a uniform strain enforces a common moir\'{e} lattice vector across all interfaces, spatially aligning the domains into a single, simply connected region.
In contrast, the actual supermoir\'{e} geometry involves constituent moir\'{e} lattice vectors that share nearly the same magnitude but differ in orientation.
This orientational mismatch induces a real-space beating pattern that modulates the single-moir\'{e} patches, explaining the subtle local differences observed, for instance, between the blue regions in stacking maps of Fig.~\ref{fig:fig7}.

\section{Other stacking families}
\label{Sec:AppendixB}

\noindent
In the following, we extend our continuum description to the $\alpha\alpha\alpha$ and $\alpha\beta\gamma$ families, defined by their local moir\'{e} stacking sequences, and summarize the corresponding effective Hamiltonians for each.
While the main text focuses on the relaxation-favored $\alpha\beta\alpha$ family, the remaining sequences also appear in the tessellation and complete the domain-resolved continuum description.
The electronic structure and valley topology of $\alpha\alpha\alpha$ and $\alpha\beta\gamma$ are discussed in Appendices~\ref{subSec:B1} and \ref{subSec:B2}, respectively.

\subsection{$\alpha\alpha\alpha$ family}
\label{subSec:B1}

\noindent
The $\alpha\alpha\alpha$ family is characterized by a local AA-like moir\'{e} stacking sequence ($\bm{d}_{\ell\ell'}=\mathbf{0}$) and generically hosts gapless Dirac nodes in its low-energy spectrum.
We begin with the $14\times14$ first-shell Hamiltonian of h3G ($\alpha\alpha$) to capture these band structure features.
The effective model near $K_{2}$ follows analogously to Ref.~\cite{Zhu2020} and is identical to $\alpha\beta$ [Eq.~\eqref{eq:H3G_K2}], where the symmetric interlayer bias does not enter at leading order.

In contrast, near $K_{1}$, where the low-energy weight is concentrated on $\mathrm{L}_{1}$, the derivation is more involved and requires an explicit downfolding.
We decompose the high-energy $Q$-subspace block in Eq.~\eqref{eq:Heff1} for $\mathrm{L}_{2}$ and $\mathrm{L}_{3}$, as $H_{QQ} = H_{QQ}^{(0)} + \delta H_{\bm{k},\Delta}$.
The unperturbed block, evaluated at $\bm{k}=0$ and $\Delta=0$, reads
\begin{equation}
\label{eq:B1}
H_{QQ}^{(0)} =
\begin{pmatrix}
\mathcal{H} & \mathcal{T}\\
\mathcal{T}^{\dagger} & -\mathcal{H}
\end{pmatrix},
\end{equation}
where the constituent diagonal and off-diagonal blocks are defined as
\begin{equation}
\label{eq:B2}
\mathcal{H} =
\begin{pmatrix}
h_{0} & 0 & 0\\
0 & h_{+} & 0\\
0 & 0 & h_{-}
\end{pmatrix},
\qquad
\mathcal{T}=
\begin{pmatrix}
0 & T^{-} & T^{+}\\
T^{-} & 0 & T^{0}\\
T^{+} & T^{0} & 0
\end{pmatrix},
\end{equation}
with $h_{n} = \hbar v_{\rm F}(\bm{q}_{n}\cdot\bm{\sigma})$.
Retaining terms only up to first order in $\varepsilon$ and $\delta H_{\bm{k},\Delta}$, we expand
\begin{align}
\label{eq:B3}
(\varepsilon-H_{QQ})^{-1}
&\simeq
-\bigl(H_{QQ}^{(0)}\bigr)^{-1}
-\varepsilon \bigl(H_{QQ}^{(0)}\bigr)^{-2}
\notag\\
&\qquad+\bigl(H_{QQ}^{(0)}\bigr)^{-1}\delta H_{\bm{k},\Delta}\bigl(H_{QQ}^{(0)}\bigr)^{-1}.
\end{align}
The inverse of the unperturbed block is obtained from the block-matrix inversion rule (Schur complement) as
\begin{equation}
\label{eq:B4}
\bigl(H_{QQ}^{(0)}\bigr)^{-1}
= \frac{1}{1-3\alpha^2}
\begin{pmatrix}
\mathcal{H}^{-1} & -3\alpha^{2}(\mathcal{T}^{\dagger})^{-1} \\
-3\alpha^{2} \mathcal{T}^{-1} & -\mathcal{H}^{-1}
\end{pmatrix}.
\end{equation}

Substituting Eqs.~\eqref{eq:B3} and \eqref{eq:B4} into the effective-Hamiltonian framework introduced in Eq.~\eqref{eq:Heff1}, the downfolding procedure yields an effective two-band Dirac model with the parameters
\begin{equation}
\label{eq:H3G_aa_K1}
\begin{gathered}
\frac{v^{(1)}}{v_{\rm F}}
= \frac{1-9\alpha^{2}+9\alpha^{4}}{1+36\alpha^{4}}, \\
\frac{\Delta_{0}^{(1)}}{\Delta} = -\frac{1-6\alpha^{2}-18\alpha^{4}}{2(1+36\alpha^{4})}, \qquad
\frac{\Delta_{z}^{(1)}}{\Delta} = 0.
\end{gathered}
\end{equation}
By $C_{2x}$ symmetry, the $K_{3}$ expansion satisfies $v^{(3)}=v^{(1)}$, $\Delta_{0}^{(3)}=-\Delta_{0}^{(1)}$, and $\Delta_{z}^{(3)}=-\Delta_{z}^{(1)}=0$.
Note that the interlayer bias does not open a gap at the Dirac nodes, since the Dirac mass term vanishes, $\Delta_{z}=0$, for all sectors in h3G ($\alpha\alpha$) to the order considered, inducing only an overall on-site energy shift via $\Delta_{0}$.

The addition of a fourth layer (h4G) expands the $K_{1}$ low-energy sector to include both $\mathrm{L}_{1}$ and $\mathrm{L}_{4}$, resulting in a twofold multiplicity of Dirac cones at $K_{1}$.
Thus, the low-energy physics is captured by the following four-band effective Hamiltonian:
\begin{align}
\label{eq:H4G_aaa_K1}
\mathcal{H}^{K_{1}}_{\alpha\alpha\alpha}
=
\begin{pmatrix} 
\Delta^{(14)}_{z} & v^{(1)}\Pi^{\dagger}_{\bm{k}} & \gamma_{1} & iv_{4}\Pi^{\dagger}_{\bm{k}} \\
v^{(1)}\Pi_{\bm{k}} & \Delta^{(14)}_{z} & -iv_{4}\Pi_{\bm{k}} & \gamma_{1} \\
\gamma_{1} & iv_{4}\Pi^{\dagger}_{\bm{k}} & -\Delta^{(14)}_{z} & v^{(1)}\Pi^{\dagger}_{\bm{k}} \\
-iv_{4}\Pi_{\bm{k}} & \gamma_{1} & v^{(1)}\Pi_{\bm{k}} & -\Delta^{(14)}_{z}
\end{pmatrix}.
\end{align}
This model is formally identical to the continuum model of AA bilayer graphene, with an effective interlayer coupling identified as
\begin{equation}
\label{eq:H4G_aaa_para}
\begin{aligned}
\frac{\Delta^{(14)}_{\mathrm{z}}}{\Delta} &= -\frac{1-4\alpha^{2}}{2(1+36\alpha^{4})} ,
\\
\frac{\gamma_{1}}{w} &= -\frac{9\alpha^{2}(1-3\alpha^{2})}{\sqrt{1+72\alpha^{4}}} ,
\\
\frac{v_{4}}{v_{\rm F}} &= \frac{18\alpha^{3}}{(1+36\alpha^{4})\sqrt{1+72\alpha^{4}}}.
\end{aligned}
\end{equation}
Diagonalizing Eq.~\eqref{eq:H4G_aaa_K1} shows that the spectrum consists of two Dirac cones centered at energies $\pm\sqrt{\gamma_{1}^{2} + \big(\Delta^{(14)}_{\mathrm{z}}\big)^{2}}$.
More generally, for an $N_{\ell}$-fold multiplicity, the low-energy effective Hamiltonian maps onto that of $N_{\ell}$-layer AA-stacked graphene.
The resulting spectrum consists of $N_{\ell}$ Dirac cones shifted vertically in energy (see the left panels in Fig.~\ref{fig:figA1}) and remains gapless even under an applied bias~\cite{Min2008a} or the inclusion of the MDT.
Consequently, the $\alpha\alpha\alpha$ family exhibits a metallic character, rendering the valley Chern number ill-defined.
The domain background vanishes identically due to $C_{2x}$ symmetry, independent of the layer number.

\subsection{$\alpha\beta\gamma$ family}
\label{subSec:B2}

\noindent
We next consider the $\alpha\beta\gamma$ family, which corresponds to a local rhombohedral-like moir\'{e} stacking sequence defined by $\bm{d}_{\ell\ell'} = -2\ell\bm{\delta}$.
For h3G, this family reduces to the $\alpha\beta$ sequence, whose effective description was discussed in Sec.~\ref{subSec:h3G} of the main text.
Upon adding a fourth layer, the electronic structure of the $\alpha\beta\gamma$ domain no longer follows a simple mapping to rhombohedral graphene.
This behavior stands in contrast to $\alpha\alpha\alpha$ and $\alpha\beta\alpha$, which map directly onto AA- and Bernal-stacked graphene, respectively.

To elucidate the origin of this deviation, we derive a four-band effective model near $K_{1}$ from the first-shell Hamiltonian and analyze the resulting low-energy dispersion and valley topology in h4G ($\alpha\beta\gamma$).
The $Q$-subspace block and its inverse retain the structure of Eqs.~\eqref{eq:B1}--\eqref{eq:B4}, but with tunneling matrices modified as $[T^n]_{\ell\ell'} \rightarrow [T^{n}]_{\ell\ell'} e^{i \bm{q}_n \cdot \bm{d}_{\ell\ell'}}$ to incorporate the local stacking vectors $\bm{d}_{\ell\ell'}$.
Downfolding onto the $\{\mathrm{1A},\mathrm{1B},\mathrm{4A},\mathrm{4B}\}$ basis leads to the following four-band Hamiltonian:
\begin{align}
\label{eq:H4G_abc_K1}
&\mathcal{H}^{K_{1}}_{\alpha\beta\gamma}
=
\begin{pmatrix} 
\Delta_{\mathrm{A}} & v^{(1)}\Pi^{\dagger}_{\bm{k}} & \gamma_{\mathrm{A}} & iv_{4}\Pi^{\dagger}_{\bm{k}} \\
v^{(1)}\Pi_{\bm{k}} & \Delta_{\mathrm{B}} & 2iv_{4}\Pi_{\bm{k}} & \gamma_{\mathrm{B}} \\
\gamma_{\mathrm{A}} & -2iv_{4}\Pi^{\dagger}_{\bm{k}} & -\Delta_{\mathrm{A}} & v^{(1)}\Pi^{\dagger}_{\bm{k}} \\
-iv_{4}\Pi_{\bm{k}} & \gamma_{\mathrm{B}} & v^{(1)}\Pi_{\bm{k}} & -\Delta_{\mathrm{B}}
\end{pmatrix},
\end{align}
where the velocity parameters $v^{(1)}$ and $v_{4}$ coincide with those in Eq.~\eqref{eq:H4G_K1para} to fourth order in $\alpha$, whereas the sublattice-resolved bias and interlayer hopping parameters are
\begin{equation}
\label{eq:H4G_abc_para}
\hspace*{-0.5em}
\begin{alignedat}{2}
\frac{\Delta_\mathrm{A}}{\Delta}
&= -\frac{1-4\alpha^{2}}{2(1+36\alpha^{4})},
&\;\:
\frac{\Delta_\mathrm{B}}{\Delta}
&= -\frac{1-4\alpha^{2}-9\alpha^{4}}{2(1+63\alpha^{4})},
\\[2pt]
\frac{\gamma_\mathrm{A}}{w}
&= -\frac{9\alpha^{2}(1-3\alpha^{2})}{\sqrt{1+72\alpha^{4}}},
&\;\:
\frac{\gamma_\mathrm{B}}{w}
&= \frac{18\alpha^{2}(1-3\alpha^{2})}{1+63\alpha^{4}}.
\end{alignedat}
\end{equation}
At the Dirac point, the spectrum splits into two sublattice sectors, labeled by $\sigma \in \{\mathrm{A}, \mathrm{B}\}$, with eigenenergies $\pm \sqrt{\gamma_{\sigma}^{2} + \Delta_{\sigma}^{2}}$.
Since $\Delta_{\mathrm{A}} \simeq \Delta_{\mathrm{B}}$ but $|\gamma_{\mathrm{A}}| \simeq |\gamma_{\mathrm{B}}|/2$ up to second order in $\alpha$,
the direct gap is set by the A sector and is given by $2\sqrt{\gamma_{\mathrm A}^{2} + \Delta_{\mathrm{A}}^{2}}$, which remains finite and increases monotonically with the interlayer bias, as shown in Fig.~\ref{fig:figA2}(c).

We introduce the symmetric/antisymmetric ($s/a$) states, $|\sigma_{s/a}\rangle=(|1\sigma\rangle\pm|4\sigma\rangle)/\sqrt{2}$, to explicitly reveal the underlying structure.
The Hamiltonian in the basis $\{|\mathrm{ A}_s\rangle, |\mathrm{B}_s\rangle, |\mathrm{ A}_a\rangle, |\mathrm{B}_a\rangle\}$ then takes the block form:
\begin{align}
\label{eq:H4G_abc_K1_sym}
\tilde{\mathcal{H}}^{K_{1}}_{\alpha\beta\gamma}
&=
\begin{pmatrix}
h_{s} & h_{\delta} \\
h_{\delta}^{\dagger} & h_{a}
\end{pmatrix},
\notag\\[2pt]
h_{s/a}
&=
\begin{pmatrix}
\pm \gamma_{\mathrm{A}} &
\left( v^{(1)} \mp \frac{i}{2}v_{4}\right)\Pi^{\dagger}_{\bm{k}} \\
\left( v^{(1)} \pm \frac{i}{2}v_{4}\right)\Pi_{\bm{k}} &
\pm \gamma_{\mathrm{B}}
\end{pmatrix},
\notag\\[2pt]
h_{\delta}
&=
\begin{pmatrix}
\Delta_{\mathrm{A}} &
-\frac{3i}{2}v_{4}\Pi^{\dagger}_{\bm{k}} \\
-\frac{3i}{2}v_{4}\Pi_{\bm{k}} &
\Delta_{\mathrm{B}}
\end{pmatrix}.
\end{align}
For $\Delta = v_{4} = 0$, the off-diagonal block $h_{\delta}$ vanishes, and the Hamiltonian becomes block diagonal.
For finite $\Delta$ and $v_{4}$, however, the symmetric and antisymmetric sectors are coupled through $h_{\delta}$.
We further downfold onto each sector, retaining only terms linear in $\bm{k}$ and discarding all contributions proportional to $\Delta^{2}$ or $v_{4}^{2}$.
This yields the renormalized forms
\begin{equation}
\tilde{h}_{s/a} =
\begin{pmatrix}
\pm \gamma_\mathrm{A} & \bigl(v^{(1)} \mp i\,\eta v_{4}\bigr)\Pi^{\dagger}_{\bm{k}} \\
\bigl(v^{(1)} \pm i\,\eta v_{4}\bigr)\Pi_{\bm{k}} & \pm \gamma_\mathrm{B}
\end{pmatrix},
\end{equation}
where
\begin{equation}
\eta
=
\frac{1}{2}
+\frac{3}{2}
\left(
\frac{\Delta_{\mathrm B}}{\gamma_{\mathrm B}}
-\frac{\Delta_{\mathrm A}}{\gamma_{\mathrm A}}
\right).
\end{equation}
At this order, the diagonal terms remain unrenormalized.

The resulting dispersions in each sector are those of a massive Dirac model with a Dirac mass term $\pm(\gamma_{\mathrm{A}}-\gamma_{\mathrm{B}})\sigma_{z}/2$.
The lowest-energy conduction and valence bands of the total spectrum originate from the antisymmetric conduction branch and the symmetric valence branch, respectively.
Remarkably, the corresponding valley Chern numbers are found to be identical, $C_{K_{1}, \mathrm{c}} = C_{K_{1}, \mathrm{v}} = -1/2$, summing to a nonzero value as a consequence of band inversion.
In the $K_{2}$ and $K_{3}$ sectors, the MDT-induced sublattice potential opens a direct gap, thereby separating the conduction and valence bands.
Crucially, the two sectors share the same mass sign and therefore make identical local contributions to the valley topology: 
$C_{K_{2},\mathrm{c}}=C_{K_{3},\mathrm{c}}=+1/2$ and $C_{K_{2},\mathrm{v}}=C_{K_{3},\mathrm{v}}=-1/2$.
Including the domain contribution of $-1/2$ for both bands gives the net valley Chern numbers $C_{\mathrm{c}} = 0$ and $C_{\mathrm{v}} = -2$, consistent with Fig.~\ref{fig:figA2}(c) and Ref.~\cite{Fujimoto2025}.
The domain background remains fixed at $-1/2$ throughout the $\alpha\beta\gamma$ family, independent of the layer number.

We extend the $\alpha\beta\gamma$ domain analysis to h5G ($\alpha\beta\gamma\alpha$) and h6G ($\alpha\beta\gamma\alpha\beta$).
The low-energy spectrum again decomposes into folded Dirac sectors governed by the multiplicity $N_{\ell}$, defined as the number of layers participating in each $K_{\ell}$ sector.
In h5G, $N_{1}=N_{2}=2$ and $N_{3}=1$: the inverted structure discussed above occurs at $K_{1}$ and $K_{2}$, whereas $K_{3}$ hosts a single massive Dirac cone.
The $K_{1}$ and $K_{2}$ sectors each contribute $-1/2$ to both the conduction and valence bands, while the remaining $K_{3}$ sector contributes $+1/2$ and $-1/2$, respectively.
Together with the domain contribution of $-1/2$, these sector contributions yield the net valley Chern numbers $C_{\mathrm{c}} = -1$ and $C_{\mathrm{v}} = -2$.
In h6G, $N_{1}=N_{2}=N_{3}=2$, so the same inverted structure appears in all three sectors.
Consequently, both bands carry $C_{\mathrm{c}} = C_{\mathrm{v}} = -2$, combining a $-1/2$ contribution from each sector with an additional $-1/2$ from the domain background.
The $\alpha\beta\gamma$ family is therefore distinguished from $\alpha\alpha\alpha$ by a robust, topologically nontrivial gap, but unlike $\alpha\beta\alpha$, it does not exhibit gate-induced gap-closing topological transitions.

\makeatletter
\def\bibsection{%
  \par\vspace{1.5em}%
  {\noindent\normalfont\large\bfseries\refname\par}%
  \vspace{0.8em}%
}
\makeatother
\providecommand{\bibfont}{}
\renewcommand{\bibfont}{\small}

\end{document}